\documentclass[conference]{IEEEtran}
\usepackage[linesnumbered,algoruled,boxed,lined]{algorithm2e}
\usepackage{algpseudocode}
\usepackage{makecell}
\usepackage{amsmath}
\usepackage[numbers]{natbib}
\usepackage{multirow}
\usepackage{booktabs}
\usepackage{hyperref}
\usepackage{diagbox}

\hypersetup{
    colorlinks=true,
    linkcolor=blue,
    filecolor=magenta,
    urlcolor=magenta,
}
\usepackage[all]{hypcap}
\usepackage{booktabs}
\usepackage{url}
\usepackage{xcolor}
\usepackage{tikz}
\usepackage{diagbox}
\usepackage{listings}
\usepackage{comment} 
\usepackage{graphicx}

\usepackage{tcolorbox}

\tcbset{
    colback=gray!10, 
    colframe=gray, 
    coltext=black, 
    boxsep=5pt, 
    arc=0mm, 
    top=0mm, bottom=0mm, 
    left=1mm, right=1mm, 
    boxrule=0.5mm,
    toptitle=1mm, bottomtitle=1mm, 
    titlerule=0mm, 
}
\definecolor{deepgreen}{rgb}{0.0, 0.4, 0.0}

\definecolor{codegreen}{rgb}{0,0.6,0}
\definecolor{codegray}{rgb}{0.5,0.5,0.5}
\definecolor{codepurple}{rgb}{0.58,0,0.82}
\definecolor{backcolour}{rgb}{0.95,0.95,0.92}

\lstdefinestyle{mystyle}{
  backgroundcolor=\color{backcolour}, commentstyle=\color{codegreen},
  keywordstyle=\color{magenta},
  numberstyle=\tiny\color{codegray},
  stringstyle=\color{codepurple},
  basicstyle=\ttfamily\footnotesize,
  breakatwhitespace=false,         
  breaklines=true,                 
  captionpos=b,                    
  keepspaces=true,                 
  numbers=left,                    
  numbersep=5pt,                  
  showspaces=false,                
  showstringspaces=false,
  showtabs=false,                  
  tabsize=2
}
\newtcolorbox[auto counter,number within=chapter]{definition}[1][]{
  enhanced,
  breakable,
  fonttitle=\scshape,
  title={Definition \thetcbcounter},
  #1
}

\usepackage{xurl}

\newcommand{\tech}{\mbox{\textsc{GenTTP}}}

\usepackage[framemethod=TikZ]{mdframed}
\definecolor{mycolor}{RGB}{194, 214, 236}

\newcounter{finding}
\newcommand{\finding}[1]{\refstepcounter{finding}
 	\vspace{1mm}
	\begin{mdframed}[linecolor=gray,roundcorner=12pt,backgroundcolor=gray!15,linewidth=3pt,innerleftmargin=10pt,innertopmargin=6pt,innerbottommargin=6pt,leftmargin=0cm,rightmargin=0cm,topline=false,bottomline=false,rightline = false]
		\textbf{Findings \arabic{finding}:} #1
	\end{mdframed}
	\vspace{0.5mm}
}

\newcounter{result}

\makeatletter
\g@addto@macro{\@algocf@init}{\SetKwInOut{Parameter}{Parameters}}
\makeatother

\DeclareRobustCommand*{\IEEEauthorrefmark}[1]{%
    \raisebox{0pt}[0pt][0pt]{\textsuperscript{\footnotesize\ensuremath{#1}}}}

\begin{document}
\begin{sloppypar}
\title{\LARGE \bf Tactics, Techniques, and Procedures (TTPs) in Interpreted Malware: A Zero-Shot Generation with Large Language Models }

\author{
\IEEEauthorblockN{
Ying Zhang\IEEEauthorrefmark{1},
Xiaoyan Zhou\IEEEauthorrefmark{1},
Hui Wen\IEEEauthorrefmark{2},
Wenjia Niu\IEEEauthorrefmark{1},
Jiqiang Liu\IEEEauthorrefmark{1},
Haining Wang\IEEEauthorrefmark{3}, and
Qiang Li\IEEEauthorrefmark{1}}
\IEEEauthorblockA{\IEEEauthorrefmark{1}Beijing JiaoTong University, China}
\IEEEauthorblockA{\IEEEauthorrefmark{2}Institute of Information Engineering, Chinese Academy of Sciences, China}
\IEEEauthorblockA{\IEEEauthorrefmark{3}Virginia Tech, USA}
}

\maketitle
\begin{abstract}

\noindent Nowadays, the open-source software (OSS) ecosystem suffers from security threats of software supply chain (SSC) attacks.    
Interpreted OSS malware plays a vital role in SSC attacks, as criminals have an arsenal of attack vectors to deceive users into installing malware and executing malicious activities.
In this paper, we introduce tactics, techniques, and procedures (TTPs) proposed by MITRE ATT\&CK into the interpreted malware analysis to characterize different phases of an attack lifecycle.
Specifically, we propose {\tech}, a zero-shot approach to extracting a TTP of an interpreted malware package.
{\tech} leverages large language models (LLMs) to automatically generate a TTP, where the input is a malicious package, and the output is a deceptive tactic and an execution tactic of attack vectors.
To validate the effectiveness of {\tech}, we collect two datasets for evaluation: a dataset with ground truth labels and a large dataset in the wild. 
Experimental results show that {\tech} can generate TTPs with high accuracy and efficiency.
To demonstrate {\tech}'s benefits, we build an LLM-based Chatbot from 3,700+ PyPI malware's TTPs.
We further conduct a quantitative analysis of malware's TTPs at a large scale. 
Our main findings include: (1) many OSS malicious packages share a  relatively stable TTP, even with the increasing emergence of malware and attack campaigns,
(2) a TTP reflects characteristics of a malware-based attack, and (3) an attacker's intent behind the malware is linked to a TTP.

\end{abstract}

\section{Introduction}

With the increasing pervasiveness of open-source software (OSS) in modern software development, the OSS ecosystem is vulnerable to the exploitation of malware packages. 
According to Sonatype's report~\cite{oss_report}, software supply chain (SSC) attacks are on the rise by 430\% because adversaries have shifted to targeting numerous victims and have collected various attack methods for achieving desirable objectives. 
One of the most recent attacks is SolarWinds~\cite{ssc_solar}, where adversaries have injected malicious code into the software's build cycle, compromising downstream applications that many organizations use.  
Interpreted malware and malicious code play a central role in SSC attacks, e.g., typosquatting~\cite{vu2020typosquatting}, code injection~\cite{korczynski2017capturing}, data exfiltration, and the negative-side influence.

Detecting and understanding the attack behaviors of OSS malware is a non-trivial task.
The research community has turned its attention to content-agnostic techniques for malware packages~\cite{ferreira2021containing, guo2023empirical, duan2020towards, korczynski2017capturing}. 
Previous works~\cite{ohm2020backstabber, ladisa2023sok} present a taxonomy of attack vectors towards OSS ecosystems. 
Those attack vectors are scattered and manually summarized by security professionals. 
The industry community has paid attention to the detection of OSS malware, and the provision of security analysis reports about malware behaviors.
Practitioners utilize reverse engineering techniques to inspect the content of malware packages and write an analysis report~\cite{ssc_fallguy, ssc_backdoor, ssc_crypt}.
The analysis report illustrates how the OSS malware achieves its objective by performing a sequence of actions, e.g., downloading and executing malicious code, stealing sensitive data, etc.
However, such a malware analysis is a manual process with a professional background, not adapting to the increasing number of new OSS malware.

In this paper, we introduce the tactics, techniques, and procedures (TTPs) for the OSS malware analysis. 
TTP proposed by MITRE ATT\&CK~\cite{att-ck} is a behavioral concept to reflect the various phases of an adversary's attack lifecycle and the usage of corresponding attack techniques.
We leverage TTPs to characterize how OSS malware uses deceptive tactics to trick users into installing the malicious package and how malware uses execution tactics to activate its malicious behaviors.
Specifically, we propose a novel approach, called \textbf{\tech}, to automatically generate TTPs of interpreted malware.
{\tech} utilizes large language models (LLMs) to extract a malware's TTP with zero-shot.
First, an LLM-based agent converts an interpreted malware package into package metadata and malware code, encapsulated into an LLM prompt.
The malware metadata contains a deceptive TTP that spoofs users for installing the malicious package.
The malware code contains an execution TTP that circumvents defense and compromises users' software.
Second, we leverage prompt engineering techniques to improve the capacity of LLMs in the TTP extraction task, including the instruction, the context, the requirement, and the chain-of-thought~\cite{wei2022chain}. 
Third, {\tech} decomposes the TTP extraction task into four subtasks for obtaining a deceptive TTP and an execution TTP.  
Overall, {\tech} has two advantages: (1) it belongs to a zero-shot mechanism without any training data; (2) it automatically generates an attack tactic for an OSS malicious package.

To validate the effectiveness of {\tech}, a reliable and high-quality dataset is needed.
So far, there is no public dataset for extracting TTPs of malware packages. 
We leverage one observation: security professionals often publish analysis reports about the malware packages in OSS ecosystems.
We propose an LLM-based agent that automatically finds these reports from online sources and recognizes TTPs based on their content.
To guarantee the accuracy and representativeness of the dataset as the ground truth, we further provide a manual inspection to ensure the correctness of the extracted TTPs. 
Our experiments based on the dataset show that {\tech} can generate TTPs with high accuracy and efficiency.

Further, we use a web crawler to collect OSS malicious packages from two sources: academic open-source datasets and commercial websites. 
In total, we have collected 5,890 OSS malware packages, covering 3 OSS ecosystems (PyPI, NPM, and RubyGems). 
We leverage {\tech} to extract 3,700+ TTPs from non-duplicated malicious PyPI packages.
Then, we build an LLM-based Chatbot as a QA application.
The Chatbot uses retrieval augmented generation (RAG) to store 3,700+ malware TTPs as external knowledge to LLMs.
In terms of Chatbot, users can obtain a practical context for characterizing an adversary's objective and malicious behaviors of an OSS malicious package, such as code obfuscation, downloading files, executing files, and exfiltrating data.

We conduct a systematic analysis of 3,700+ malware TTPs in the wild, providing unique insights into defeating malware and revealing adversaries' intentions.
The main findings are summarized below.
(1) We observe that TTPs show the transitive states of various attack vectors among malware packages in real-world scenarios. For instance, there is a more than half probability (0.54) that an imitated version follows the same typosquatting behavior. 
(2) We also observe that the different malicious packages share the same TTP, which can be a strong indicator of malicious activities.
Surprisingly, up to 19.2\% of malware packages are signed the same TTP.
(3) Finally, we observe that a TTP reflects the attacker's intent and the malware patterns, such as trojan download or remote access.

The main contribution can be summarized as follows.
\begin{itemize}    
    \item We propose {\tech} to automatically generate TTPs of OSS malware. 
   \item We leverage {\tech} to obtain 3700+ TTPs for malware analysis at a large scale.
   We release the source code and TTP's artifact to the community\footnote{Artifact and Code in anonymous submissions: \url{https://github.com/sure17/GenTTP}}.
   \item We build an LLM-based Chatbot\footnote{Chatbot Demo in anonymous submissions: \url{https://genttp.streamlit.app/}} that provides malware analysis to users. 
   \item We provide a quantitative analysis of OSS malware's TTPs in the wild.   
\end{itemize}

The rest of the paper is organized as follows. 
Section 2 presents the background of malware analysis. 
Section 3 illustrates the OSS malware TTP.
Section 4 details the design of {\tech}. 
Section 5 presents the data collection for a dataset with the ground truth label and a large dataset in the wild.
Section 6 presents the evaluation of {\tech}.
Section 7 presents an LLM Chatbot and an insightful study of TTPs of malware packages in the wild.
Section 8 discusses the limitations of {\tech}.
Section 9 presents the related work, and Section 10 concludes.

\section{Background of Malware Analysis}
\label{sec:back}
 
\begin{figure}[!t]
    \centering
    \includegraphics[width=3.0in]{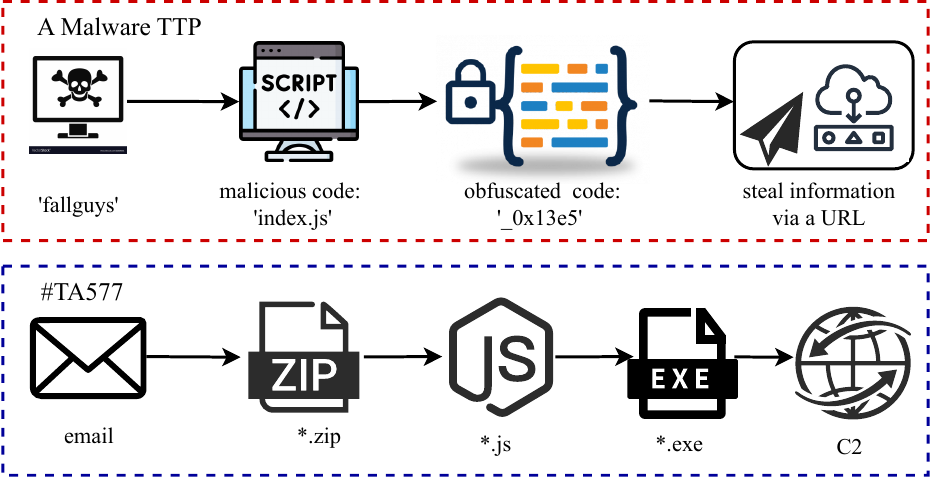}
    \caption{A comparison example. (1) The red box~\cite{ssc_fallguy}: behaviors of the malware `fallguys' in the NPM ecosystem are \{`index.js' $\rightarrow$ `obfuscation code' $\rightarrow$ steal data via a URL\}. (2) The blue box~\cite{ta577}: 
    the attack methods \#TA577 in enterprise systems are \{email $\rightarrow$*.zip $\rightarrow$*.js $\rightarrow$*.exe$\rightarrow$C2\}.}
    \label{fig:back-rp}
\end{figure}

OSS malware analysis has attracted great interest in the cybersecurity community.
Security practitioners have been publishing analysis reports to share their findings on malware packages. 
Figure~\ref{fig:back-rp} (the red box) depicts one example of a malware analysis report~\cite{ssc_fallguy}. 
A malware package, `fallguys,' stole local files and uploaded them to a third-party Discord server, e.g., via webhooks.
It used an embedded malicious code in the `index.js', where a constant `\_0x13e5' stores long strings that are obfuscated by base64.
Once decoded, its strings were converted into the execution code, which is used to steal information from local files. 
Then, `fallguys' used a send function to upload the data to a Discord URL.
Thus, `fallguys' had behaviors as \{`index.js' $\rightarrow$`obfuscation code'$\rightarrow$steal data via a discord URL\}.
In short, malware analysis can help security professionals detect and prevent malicious behaviors present in OSS packages.

Similarly, the cybersecurity industry uses TTP to present how an adversary achieves its objective by performing a sequence of actions.
TTP is the most popular attack analysis from the MITRE ATT\&CK framework~\cite{att-ck}.
Figure~\ref{fig:back-rp} (the blue box) depicts the attack tactic \#TA577~\cite{ta577}, denoted as \{email$\rightarrow$*.zip $\rightarrow$*.js $\rightarrow$*.exe $\rightarrow$ C2\}.
\#TA577 initially compromised the victim via emails containing malicious attachments and executed a `.js' file to download a `.exe' file, then built an encrypted communication with a command and control (C2) server.
TTPs are used to analyze and understand the adversary’s intent, helping defenders better identify, prevent, and respond to threats.

In this work, we leverage TTPs to represent the OSS malware analysis.
A comparison illustration is shown in Figure~\ref{fig:back-rp}, where an OSS malware TTP uses a similar sequence to conventional TTP, where each element represents an individual action or behavior in the malware.
Malware TTP provides deep behavioral analysis and identifies malicious actions of the malware. 
However, the analysis report for malware TTP suffers three limitations in practice. 
(1) Writing such a report is time-consuming and labor-intensive, which is not scalable for continuously emerging malware.  
Security professionals have to perform a combination of static and dynamic analysis tools to find a malware TTP.
Writing such a report requires a diverse skill set for security professionals, usually gained over time through technical experience and self-training.
(2) Not all malware packages are analyzed by security professionals.
In our preliminary study, we have collected 1,366 malware analysis reports (detailed in Table~\ref{tab:report:source:detailed} in Appendix).
From our observations in the dataset, about 90.3\% of malware packages are not analyzed in the security reports (detailed in Sec.~\ref{sec:sub:val}).
(3) Analysis reports for outdated malware are relatively simple, while reports for newly emerged malware are more complete and in-depth. 
For instance, the report~\cite{SSC2017} briefly described a typosquatting attack from 39 malicious packages in July 2017, yet there is a lack of behavioral and malicious functionality analysis. 
In contrast, the report~\cite{SSC_Ledger} in Dec. 2023 contained a detailed analysis of how `Ledger' utilized malicious code to conduct a phishing attack.

\section{OSS Malware TTP}

\subsection{Definition} 
A malware TTP is a sequence of attack vectors that malware packages perform to achieve a malicious objective.
Specifically, an attack vector is a method that cybercriminals use to breach or infiltrate a victim's network or system. The attack vector is a basic unit representing an element in a TTP of a malicious package.
A malware package's attack vectors can be extracted from the metadata and malware code.
\begin{equation}\label{equ:av}
    Pkg = \{AV_1, AV_2, \dots, AV_k\},
\end{equation}
where $Pkg$ denotes a malware package, and $AV_i$ represents the $i$-th attack vector in the package.
A TTP is a sequence of attack vectors, denoted as:
\begin{equation}\label{equ:ttp}
    TTP = \{AV_1 \rightarrow AV_2 \rightarrow \dots \rightarrow AV_k\},
\end{equation}
where the transferring symbol ($\rightarrow$) represents the sequence relationship between two vectors, denoted as $AV_i \rightarrow AV_j$, that the left part is the prior $AV_i$, and the right part is the following $AV_j$. 
We aim to generate TTPs for malware packages in OSS ecosystems automatically.

\begin{figure}[!t]
    \centering
    \includegraphics[width= 2.7in]{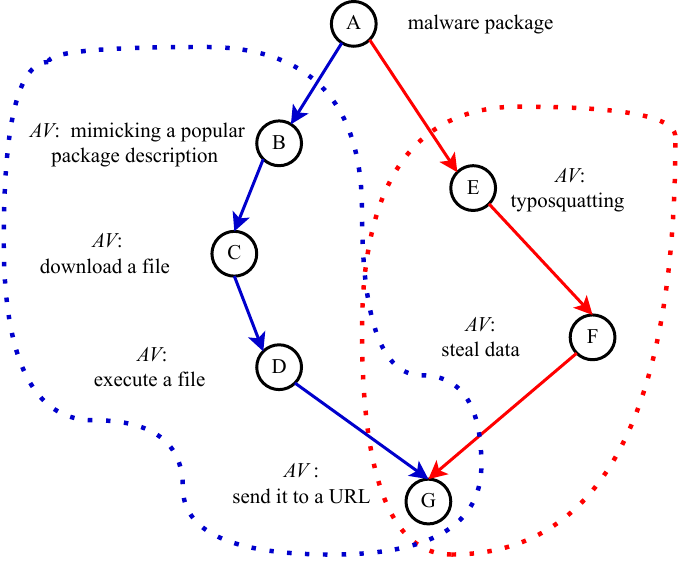} 
    \caption{One example for illustrating the attack vector chain of two malware packages. The blue color from the malware `Ascii2text' represents as \{fake description$\rightarrow$download$\rightarrow$ executes codes for local passwords$\rightarrow$send data to a web URL\}.
    The red color from the malware `loglib-modules' represents as \{typosquatting$\rightarrow$steal data$\rightarrow$send data to a web URL\}. }
    \label{fig:intro}
    \vspace{-3MM}
\end{figure} 

\textbf{Take one example.} 
Figure~\ref{fig:intro} shows two TTPs of two malware packages, `Ascii2text' and `loglib-modules'.
The malware package `Ascii2text' (node A) mimics the popular legitimate `art' package by its description, which copies the entire project description without the release part to deceive users into downloading it. 
The malicious code in `Ascii2text' is to download (node C) and execute a malicious script (node D) that searches for local passwords and uploads them using a Discord web URL (node G). 
The malware package `loglib-modules' (node A) uses typosquatting to mimic the legitimate package name `loglib' (node E), where its malicious code is to steal data (node F) and send them to a website URL (node G).
Consequently, analyzing the TTPs of malware packages can expose similar behaviors and artifacts, such as obfuscation code hiding a download activity.

\subsection{TTP Category}
\label{sec:sub:ttp}

Based on an in-depth survey we conducted on SSC attacks and OSS malware
\cite{guo2023empirical, duan2020towards, ohm2020backstabber, ladisa2023sok, decan2017empirical, zimmermann2019small},  
we propose categorizing malware TTP into two groups: deceptive TTP and execution TTP.

\textbf{Deceptive TTP} is to deceive users into downloading and installing malicious packages from an OSS registry. 
A malicious package exploits a deceptive TTP to look like a benign package.
Specifically, a deceptive TTP contains six basic attack vectors as listed in Table~\ref{tab:av}. 
(1) Typosquatting $AV$ denotes that a malicious package uses a name similar to a benign package.
(2) Imitated-version $AV$ denotes that a malicious package uses a standard version number as a legitimate package.
(3) Fake-description $AV$ denotes that a malicious package uses similar content to imitate a legitimate package. 
(4) Imitated-URL $AV$ denotes that a malicious package uses a legitimate homepage URL. 
(5) Malicious-dependency $AV$ denotes that a malicious package uses another malware as its dependency library.
(6) Fake-contact $AV$ denotes that a malicious package uses a fake contact. 

\begin{table}[!t] \small
    \centering
    \caption{TTP category: basic attack vectors and neutral actions in a malicious package. }
    \label{tab:av}
    \begin{tabular}{l |  c | c  }
    \hline    
                          \multicolumn{2}{c|}{Category}  & Term       \\ \hline  
    \multirow{3}{*}{\makecell[l]{Deceptive \\ TTP}}    &  \makecell[c]{Attack \\ Vector}  & \makecell[r]{\{Typosquatting,  Imitated-version \\   Fake-description, Imitated-URL \\ Malicious-dependency, Fake-contact\}}           \\ \cline{2-3} 
     \multirow{3}{*}{\makecell[l]{Exection \\ TTP}}    &  \makecell[c]{Attack \\ Vector}  & \makecell[r]{\{Evasion, Conceal \\ Code-Execution, Malicious URL \}}           \\ \cline{2-3} 
       & \makecell[c]{Neutral \\ Action} & \makecell[r]{\{File, Network, Process \}}      \\
\hline
\end{tabular}%
\end{table}

\begin{figure*}[!t]
    \centering
    \includegraphics[width=5.9in]{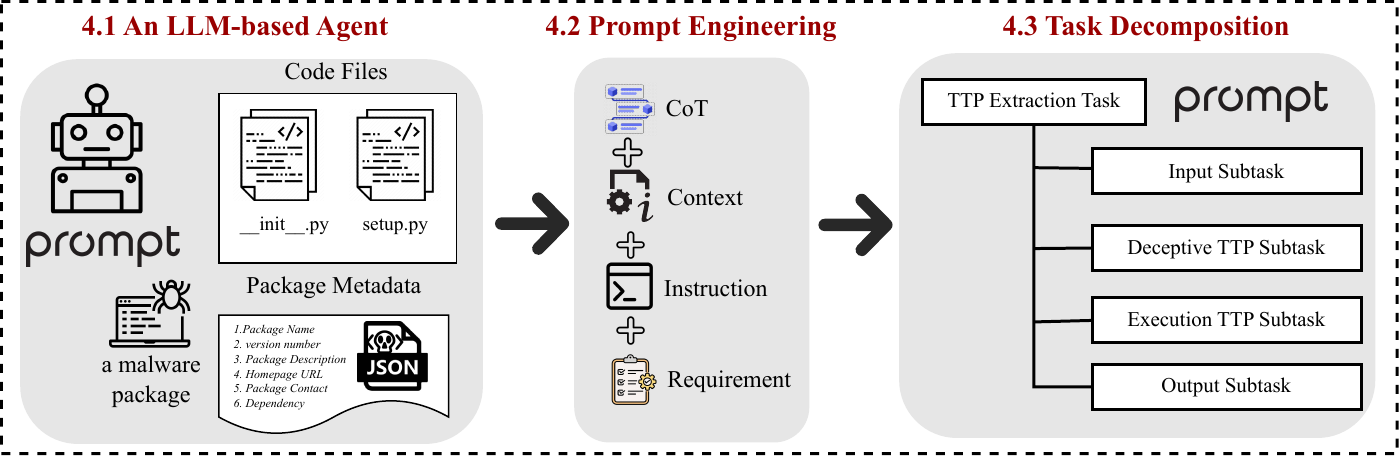}
    \caption{The architecture of {\tech}: (1) an LLM-based agent obtains the package metadata and malware code; (2) prompt engineering is used in an LLM prompt for the TTP extraction task; (3) task decomposition contains four subtasks for generating deceptive and execution TTPs.}
    \label{fig:arch}
\end{figure*}

Given a malware package, its deceptive TTP is a sequence of attack vectors, denoted as:
\begin{equation}\label{equ:ttp:d}
    TTP_{D} = \{AV_1 \rightarrow \dots \rightarrow AV_i\ \vert AV_i\in D, i\leq 6\},
\end{equation}
where $D$ is the set of deceptive attack vectors and $AV_i$ is an attack vector $i$ that is used by the malware.  
The sequence can be represented as a deceiving tactic that an attacker uses to deceive users into downloading the malware package from the official OSS registry.

\textbf{Execution TTP} runs a sequence of functions or actions to achieve the malicious objective. 
We use an execution TTP to characterize malicious behaviors that cause damage to a system or exfiltrate data.
Table~\ref{tab:av} lists four basic attack vectors in execution TTP.
(1) Evasion $AV$ denotes that malware uses obfuscated code to evade the detection techniques, e.g., base64 or zlib.
(2) Conceal $AV$ denotes that malware hides its execution code, such as steganography in which the code is hidden in an image.
(3) Code-execution $AV$ denotes that malware executes an OS command to achieve its objectives, e.g., exec and eval. 
This attack vector can execute a malicious payload. 
(4) Malicious-URL $AV$ denotes that malware has a malicious URL in its package.

Further, an execution TTP contains a series of neutral functions and actions listed in Table~\ref{tab:av}. 
Different from benign packages, malware leverages neutral functions (URLs, IPs, and files) to perform malicious activities. 
Execution TTP contains three categories of neutral actions: network, file, and process. 
(1) File action denotes that malware performs file/directory-related operations, such as open, close, read, and write.  
Malware may exfiltrate system information and environment variables, such as AWS access keys, passwords, and tokens. A package reads and exfiltrates data from the local system.
(2) Network action denotes that malware sends and receives data from a remote server, such as sending sensitive data to the remote server. 
(3) Process action denotes that malware often executes code before, during, and after the installation of a package. 
We use pre-install, install, and post-install actions to preset script running operations in the execution process. 
For example,  an \_\_init\_\_.py file may contain an execution code at installation time.

Given a malware package, its execution TTP is the sequence of attack vectors, denoted as:
\begin{equation}\label{equ:ttp:e}
    TTP_{E} = \{AV_1 \rightarrow \dots \rightarrow AV_i\ \vert AV_i\in E\},
\end{equation}
where $E$ is the set of attack vectors or neutral actions the malware uses. 
This sequence presents the execution tactic that an attacker uses to conduct malicious behaviors in the malicious package, such as circumventing the defense and compromising users' systems.

\section{Automatic TTP Generation}

Our goal is to automatically generate deceptive and execution TTPs of a malicious package in OSS ecosystems. 
A TTP contains (1) a set of attack vectors (Equation~\ref{equ:av}) and (2) a specific execution order of attack vectors (Equation~\ref{equ:ttp}).  
A malicious package often uses a TTP to deceive users and compromise users' systems.
However, there are two technical challenges in practice, which are listed below.
\begin{itemize}
    \item Extracting attack vectors from a malicious package requires static and dynamic analysis, which is currently a manual process, making it difficult to keep up to date with the numerous new malware packages.
    \item Identifying an execution order of attack vectors is a non-trivial task because malware packages have different execution sequences in their code. 
    So far, there has been no prior work to extract execution flows of attack vectors in malware packages.
\end{itemize}

\textbf{Our Motivation}. 
To address those two technical challenges, we leverage the large language model (LLM) to generate malware packages' TTPs without any training dataset.
LLMs offer a promising alternative, as recent advances demonstrate remarkable capabilities in natural language text and source code.

\textbf{Architecture}.
We propose a zero-shot framework, called {\tech}, to automatically generate TTPs of malware packages. 
The design rationale behind {\tech} is that the LLM can simulate the interpreter by generating the expected output of malware code.
Thus, there is no need to run a malware package in a virtual environment (e.g., Sandbox) to execute its code for extracting its behaviors. 
Figure~\ref{fig:arch} depicts the overall architecture of {\tech}, including three main components: (1) an LLM-based agent, (2) prompt engineering, and (3) task decomposition.
An LLM-based agent extracts package metadata and code from an interpreted malicious package. 
Prompt engineering is used to help LLM generate a malware TTP.
The task decomposition is to break the TTP generation into four simple and clear subtasks. 
A simple prompt for {\tech} is listed in Table~\ref{tab:prompt} (Appendix).
In the following, we present the detailed design of the three components.

\subsection{Agent: Malware Content Extraction}
\label{sec:sub:malware}

\begin{figure}
    \centering
    \includegraphics[width=3.2in ]{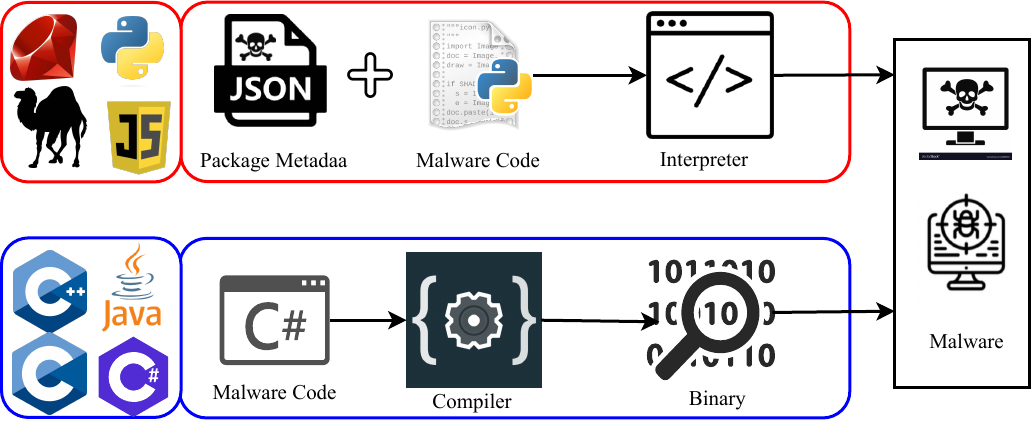}
    \caption{The red box illustrates the interpreted malware; the blue box illustrates the compiled malware.}
    \label{fig:malware}
\end{figure}

{\tech} centers on interpreted malware rather than compiled malware. 
Specifically, the malware can be divided into two categories: interpreted and compiled.
Figure~\ref{fig:malware} shows the difference between interpreted and compiled malware.
Compiled malware uses a binary format, and interpreted malware uses a software bill of materials (SBOM) to organize the components and build a software artifact, including source code, metadata, configuration, and dependencies.
Interpreted malware is similar to a regular package in OSS ecosystems, which can be unpacked and installed. 
Unlike compiled malware, interpreted malware plays a central role in SSC attacks. 
{\tech} leverages an LLM-based agent to extract malware content from an interpreted package.

\textbf{Agent} is an LLM-powered autonomous agent system that calls external APIs to retrieve information missing from LLMs.
Note that LLMs cannot directly interact with a malicious package. 
As such, we need to equip agents with tools for extracting malware content from a package. 
We seed the agent with a Shell tool for basic functionalities, e.g., decompressing and manipulating files. 
Specifically, the agent decompresses a malicious package and stores corresponding files in a directory. 
A simple prompt is used in an LLM-based agent, as shown in Table~\ref{tab:genttp:agent} (Appendix).
The agent's task is to extract the malware content from a malicious package, where the input is a package and the output is the package metadata and malware code.

\textbf{Package metadata} is used to find and install software packages from the official OSS registry.  
The metadata constitutes descriptive information providing additional context, characteristics, or attributes about a package, assisting users in comprehending its target, functionality, dependencies, and other relevant information.
Specifically, the package metadata is stored in a configuration file. 
For instance, malware in NPM uses `package.json', and malware in PyPI uses `PKG-INFO,' respectively, as its metadata file.  
The package metadata in a malware package is used to extract attack vectors.

\textbf{Malware Code} is the source code of a package designed to cause damage, security breaches, or other threats to application security. 
Specifically, a malicious package executes its code to compromise computer systems, such as stealing sensitive data, downloading trojans, and performing crypto jacking. 
Thus, malware code is the root cause of system damage.
We leverage the source code files as the prompt content for extracting an execution TTP.
For example, the source code files for a software package from the PyPI ecosystem have the extension `*.py' while those for a package from the NPM ecosystem have the extension `*.js'. 
First, the agent loads the source code into the prompt.
Then, source code files are separate documents that LLMs need to execute to obtain the malware package's behaviors.

\subsection{Zero-Shot Prompt Engineering}
\label{sec:sub:prompt}

Prompt engineering is a technique that improves the capacity of LLMs for a wide variety of applications and use cases.
{\tech} can generate deceptive and execution TTPs of a malicious package without any training dataset.
We use prompt engineering to design robust and effective prompt content in {\tech}: (1) the instruction, (2) the context, (3) the requirements, and (4) the chain of thought (CoT)~\cite{wei2022chain}.

\textbf{Instruction} presents a specific action for LLMs to perform. 
Here, we provide an instruction for each subtask in {\tech} (detailed Section~\ref{sec:sub:task}).
We attempt various instructions via an iterative process and pick a suitable one that works in practice.  
Note that one subtask only has one instruction because LLMs need to closely align with the TTP extraction task.

\textbf{Context} is external knowledge that steers LLMs to finish a specific task. 
Table~\ref{tab:context} (Appendix) lists the basic terms that LLMs use to extract malware TTPs.
Note that these terms are assigned to deceptive and execution TTPs in Table~\ref{tab:av}.
Each term has a clear and concise description.

\textbf{Requirement} is a guideline in an LLM prompt for obtaining the intended output.
{\tech} has four requirements for LLMs: (1) ensuring a thorough examination of the context, (2) outputting the results at each subtask, (3) ensuring alignment with the context of each subtask, and (4) clarity summary in each subtask.

\textbf{CoT} is used to enhance model performance on complex tasks. 
LLMs are instructed to `think step by step' to utilize more test-time computation to decompose a task into smaller and simpler steps. 
CoT transforms tasks into multiple manageable steps and sheds light on an interpretation of the model's thinking process.
LLM can do a task with simple prompting like ``Let's think step by step''.

\subsection{Task Decomposition} 
\label{sec:sub:task}

The TTP extraction task is complicated, and we need to break it down into smaller and simpler subtasks for LLMs.
Specifically, the TTP extraction task involves (1) inputting the metadata and malware code, (2) generating deceptive attack vectors from package metadata, (3) generating execution attack vectors from code execution sequences in source code, and (4) outputting attack vectors and TTPs. 
Note that each subtask uses similar prompt engineering techniques.  
In this work, we consider Python programming language, but the approach is general to any interpreter.

\textbf{Input Subtask} systematically finds the package metadata and source code files in the directory. 
For the metadata, the subtask directly reads a configuration file and adds its content to the prompt.
For the malware code, the subtask locates all source code files (e.g., *.py) in the directory and adds their content to the prompt. 
Note that the order in which the files are read is not important to LLMs.
We find that today's LLMs can automatically generate the order in which source code files are executed.

\textbf{Deceptive Subtask} is to obtain a deceiving sequence of attack vectors from the package metadata step by step.
Specifically, a deceptive TTP makes a malware package look like a benign package, and the metadata is used to deceive software developers into installing the package.
This subtask is simple, where LLM uses the context to determine whether the package metadata contains an attack vector. 
For example, LLMs determine a typosquatting attack vector by querying the OSS registry: if a malware name is similar to a legitimate package's name, {\tech} put it into the set of deceptive attack vectors.

\begin{figure}
    \centering
    \includegraphics[width=3.1in ]{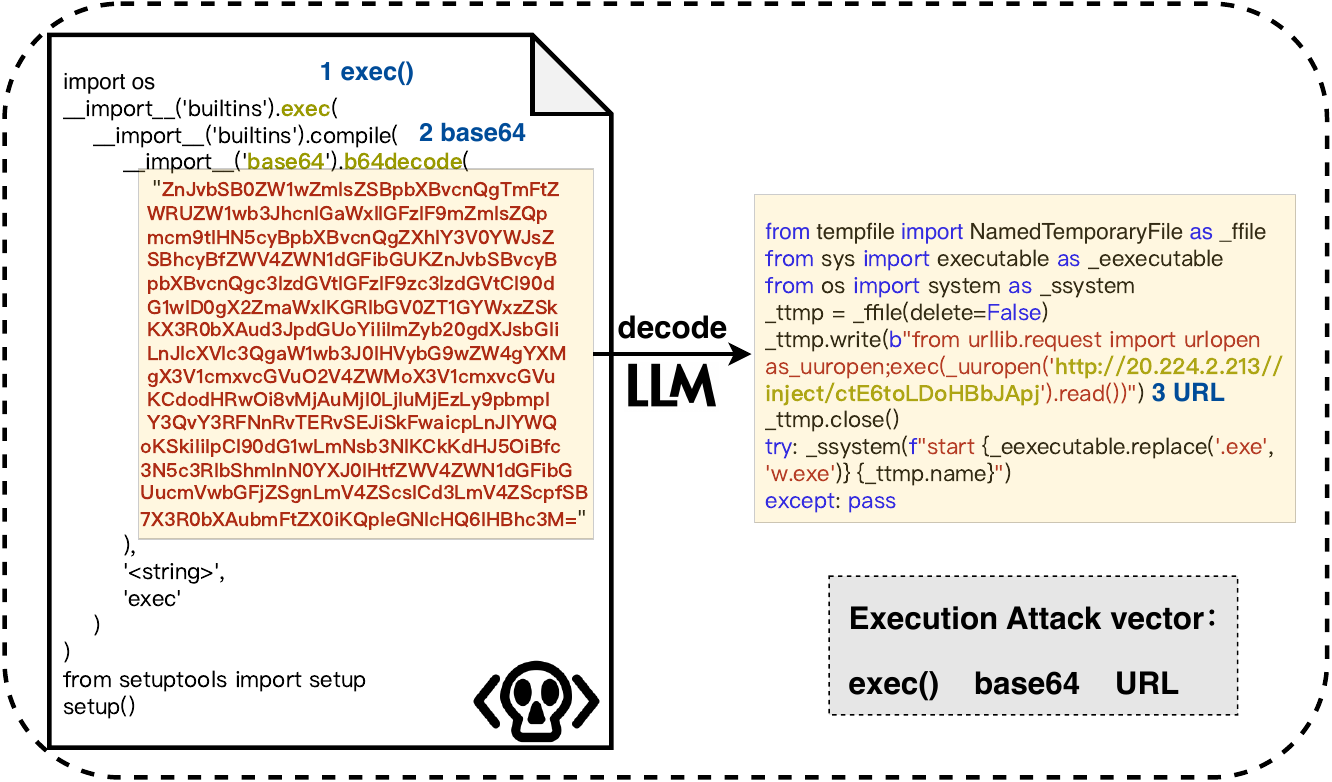}
    \caption{An example for execution subtask: the malicious code comes from the malware `Colorama'.}
    \label{fig:subtask:code}
\end{figure}

\textbf{Execution Subtask} is to generate an execution sequence of attack vectors from the malware code. 
LLM's interpreter can virtually run malware code that is the root cause of the activation of malicious behavior. 
In other words, LLM simulates the interpreter by generating the expected output of the source code without having to execute the malware code. 
If the code executes successfully, the program status is updated, and execution continues. If the code is not executable or throws any exceptions, LLM is used to simulate execution. 
This subtask outputs the result of source code execution. 
Figure~\ref{fig:subtask:code} depicts the process of this subtask via one malicious package (`Colorama').
The malware code is encoded by base64, which is an obfuscation technique. 
Note that the original source code is converted into ASCII string format, making it difficult to read and understand.
In this case, the LLM interpreter can decode the base64 format into its source code.

After running the source code, the execution subtask finds typical malicious actions that a malicious software package can perform, resulting in damage to the system or leaking data. 
By leveraging the context, the subtask identifies and extracts malicious behavior that occurs as a result of simulated code execution. 
If an action appears as a result of the malware code execution, {\tech} put it into the set of execution attack vectors.

\begin{table}[!t] \small
      \caption{The example (`coloram'): an abstract TTP and a detailed TTP. }
      \label{tab:example}
      \centering
     \resizebox{\linewidth}{!}{
      \begin{tabular}{  c c  }
      \hline
                               &  Content   \\
      \hline
      \makecell[c]{abstract\\ TTP}   & \makecell[r]{ \{typosquatting$\rightarrow$imitated version$\rightarrow$fake description \\ $\rightarrow$fake contact $\rightarrow$cmd$\rightarrow$evasion$\rightarrow$url/ip/port \} }
      \\ \hline
      \makecell[c]{detailed\\TTP} &   \makecell[r]{ \{`Colorama'$\rightarrow$`0.2.7'$\rightarrow$`Official Stanford Karel library used \\in CS 106A'$\rightarrow$`tyep@XXX.XX'$\rightarrow$`exec()'$\rightarrow$base64$\rightarrow$ 
      \\ `http://20.224.2.213//inject/ctE6toLDoHBbJApj'\} }  \\
       \hline
      \end{tabular}
      }
     \vspace{-4MM}
\end{table}

\begin{figure*}[!t]
    \centering
    \includegraphics[width=5.7in]{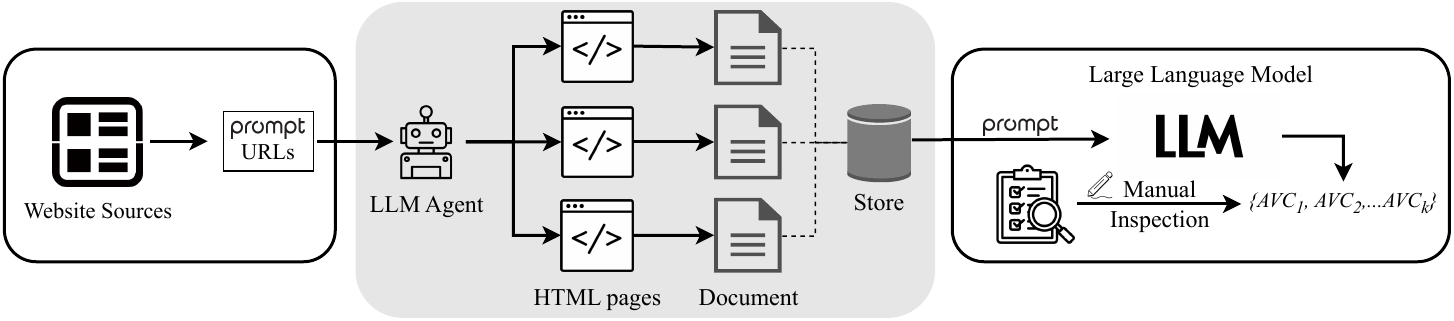}
    \caption{The data collection with ground truth labels. (1) A website URL list is used to collect the security analysis reports from the Internet. (2) An LLM agent crawls HTML webpages and determines whether they belong to analysis reports. (3) LLM outputs the TTPs of the reports and manual assistant to check the TTPs}
    \label{fig:data}
\end{figure*}

\textbf{Output Subtask.} 
{\tech} outputs a TTP format rather than a textual format.
We convert a string into two structures as Equations~\ref{equ:ttp:d} and \ref{equ:ttp:e}. 
The terminology used in the output results is required to keep consistent with the context in Table~\ref{tab:av}.
Specifically, the output subtask leverages the results of the prior two subtasks to generate malware TTP. 
A TTP starts with the deception attack vector, followed by the execution attack vector, connected by the right arrow ($\rightarrow$).
In addition, each TTP has two formats: an abstract TTP and a detailed TTP. 
The abstract TTP only contains names of attack vectors, and a detailed TTP contains specific content of attack vectors.  
Table~\ref{tab:example} lists a PyPI malware `Coloram' with an abstract TTP and a detailed TTP generated by {\tech}.
Those TTPs can be represented as an execution tactic that an attacker uses to conduct malicious behaviors in the packages to circumvent the defense and compromise users' software.


\section{Data Collection}
\label{sec:data}

In this section, we present the collection of two datasets to validate the effectiveness of {\tech}, including a large-scale dataset of malware packages and a dataset with ground-truth.

\subsection{A Large-scale Malware Dataset}
\label{sec:sub:mal}

The large-scale dataset in the wild consists of OSS malware packages scattered over the Internet.
To collect malware packages, we conducted an in-depth study of relevant literature and commercial websites about the OSS ecosystem.
Typically, malware packages come from two sources: academia and the industry. 
In academia, we found three existing open source datasets about malware packages in OSS ecosystems, including Backstabber-Knife~\cite{backstabbers-online}, Maloss~\cite{maloss, duan2020towards}, and Mal-PyPI~\cite{guo2023empirical}. 
In the industry, we found that six commercial websites provide malware packages in the OSS ecosystem, including GitHub Advisory~\cite{github-advisory}, Snyk.io~\cite{snyk-online}, Tianwen~\cite{tianwen-online}, DataDog~\cite{datadog}, Phylum~\cite{phylum-online}, and Socket~\cite{socket-online}.
Note that those sources are scattered, and we have to take a manual approach to collecting malicious packages.

\begin{table}[!t] \small
	\caption{The dataset of malicious software packages. }
	\label{tab:dataset:malware}
	\centering
	\begin{tabular}{  c c c  }
	  \hline
						  &  Language  & Malicious Package  \#  \\
	\hline
	Rubygems    		& Ruby 		 &  54     \\
	NPM    				& JavaScript  &   2,101     \\
	PyPI			  & 		Python   & 3,735     \\ 
	\hline  
	\end{tabular}
  \vspace{-4MM}
\end{table}

There are three observations from our data collection for malware packages in the wild. 
(1) The dataset from academia has a low update frequency and the dataset from the industry has a high update frequency. 
For example, the Mal-PyPI dataset~\cite{guo2023empirical} has almost no updates after its first publication. 
(2) Some sources provide the malware packages (DataDog and Maloss), and some sources do not provide any malware packages (Backstabber-Knife and all commercial companies). 
Thus, we use the web crawler to collect the malware package names from sources that are unwilling to provide data.
We use malware package names and versions as keywords to search the package registry or relevant registry mirrors.
If the malware is found, we can collect its related information; otherwise, it is labeled as missing.
Specifically, we use 3 NPM mirrors~\cite{npm_cnpmjs, npm_aliyun, npm_ustc}, 3 PyPI mirrors~\cite{pypi_tsinghua, pypi_aliyun, pypi_ustc}, and 3 RubyGems mirrors~\cite{ruby_tsinghua, ruby_hust, ruby_sysu} to search for malicious packages.
(3) There are many missing malware packages in the dataset from academia and the industry. 
The reasons are as follows: some datasets are unwilling to provide malware data (e.g., Backstabber-Knife~\cite{backstabbers-online}); some datasets only provide a part of malware packages (e.g., Mal-PyPI dataset~\cite{guo2023empirical}); and the industry does not provide any malware package. 
In total, we have collected 5,890 interpreted malware packages covering three OSS ecosystems, the largest OSS malware dataset listed in Table~\ref{tab:dataset:malware}.

We believe that a high-quality and complete OSS malware dataset is an important source for the research community.
We share the dataset via a private GitHub repo, and applicants need to send an email request to join this repo.  With the availability of the malware dataset we provided, researchers in this field will avoid the burden of OSS malware collection in the future.

\begin{table}[!t] \small
    \centering
    \caption{Source of security analysis reports (detailed in Table~\ref{tab:report:source:detailed} in Appendix). }
    \label{tab:report:source}
    \begin{tabular}{c c   c   }
    \hline
    Category   &   Website   \#          & Report  \#            \\ \hline
    Technical Community    &      16       &  516      \\
    Commercial org.  &      15             &  545    \\
    News        &  4        &  143                   \\
    Individual   &      3             & 95                     \\
    Official    &     1              & 24            \\
    Other   &       29           & 43             \\  \hline
    Total & 68 & 1,366                   \\
   \hline
    \end{tabular}%
     \vspace{-4MM}
\end{table}

\subsection{OSS Malware DataSet with Ground-Truth}
\label{sec:sub:report}

A reliable dataset with ground-truth is essential to evaluate {\tech}'s efficacy.  
A recent malware platform is CISA~\cite{next-gen}, where malware analysis reports are written in PDF and STIX-2.1 data formats.
Another collection of analysis is \cite{apt}, where popular APTs and malware analysis reports are organized in PDF.
However, we have investigated those sources and found that none of those reports are related to OSS malware. 
In short, there is no public dataset with ground-truth for the interpreted malware TTP. 

To address the lack of the dataset with ground-truth, we leverage one observation: malware analysis reports contain the ground-truth labels of malware packages' TTPs.
Security professionals often publish security analysis reports about malware packages on the Internet. 
Those analysis reports are important sources for extracting TTPs of malware packages in OSS ecosystems.
However, there are two practical issues for collecting the dataset with ground-truth. 
\begin{itemize}
    \item Security analysis reports are scattered on the Internet, and many reports are irrelevant to malware packages in OSS ecosystems. Finding them is a manual process and takes a long time.
    \item TTPs or attack vectors are written in natural language format, and it is costly to extract them from the security analysis reports.
\end{itemize}
We propose an LLM-powered agent to reduce manual efforts / time. 
Figure~\ref{fig:data} depicts the overall data collection with the malware TTPs: a website URL list, an LLM-based agent, and a manual assistant.

\textbf{Source URL}.
We focus on the security analysis reports of malware packages available on the Internet. 
Through a deep investigation, we have found 68 websites that provide security analysis reports about malware packages. 
Table~\ref{tab:report:source} lists the online sources of websites, where we use the URL list to collect reports from the Internet.
Security professionals in the industry have organized these reports, and they are reliable and trustworthy, covering many other irrelevant topics (e.g., web application vulnerabilities, software vulnerabilities, and hardware vulnerabilities). 
We use each URL as a part of the prompt and send it to the LLM. 
From these sources, LLM selects those related to malware analysis reports in OSS ecosystems.

\textbf{Agent}. 
Our agent uses a web crawler as the external API and an HTML parser to extract the main content of each webpage. 
The agent is to equip LLMs with external knowledge (webpage crawler and extraction) to extend its capabilities.
Given a webpage, LLM determines whether it contains a security analysis report about malware packages in OSS ecosystems.
The LLM-based agent extracts the main content of the webpage and removes irrelevant content, such as advertisements, pictures, dynamical scripts, and navigation bars.
Over the documents collected by the LLM-based agent, each HTML webpage is converted into a text document for extracting TTPs or attack vectors.
A simple prompt used in the LLM-based agent is detailed in Table~\ref{tab:prompt:report} (Appendix).

\textbf{Manual Assistant}.
However, there are two practical problems of seeking TTPs from security analysis reports.
(1) A security report may contain multiple malware packages, leading to dozens of attack vectors or TTPs. 
The mapping between a security report and malware packages is a one-to-many relationship rather than a one-to-one relationship. 
This is because security professionals write the analysis for several malware packages rather than a single malware. 
(2) LLM may output false positives of malware TTPs. 
Therefore, we apply a manual assistant to check the TTPs or attack vectors outputted by the LLM-based agent.
If a security analysis report contains one TTP and one malware package, the manual assistant will check whether it is a true or false positive.
If a security analysis report contains multiple packages and TTPs, the manual assistant divides those packages and provides a malware code inspection.
Manual assistance is necessary because the dataset lacks the ground truth labels.
In total, we collected 1,366 security analysis reports about malware packages and the detailed information is listed in Table~\ref{tab:report:source:detailed} in Appendix.

\subsection{Data Validation}
\label{sec:sub:val}

To guarantee the reproducibility and transparency of our datasets, we use the requirements of previous work~\cite{rossow2012prudent} to calibrate malware packages and analysis reports manually.

\textbf{Dataset Coverage}. 
The large-scale malware dataset covers all the malware mentioned in OSS malware analysis reports. In other words, the malware dataset with ground truth is a subset of the large-scale malware dataset. 
An analysis report reveals malware information: when or how to find the malicious package, its malicious behaviors or objectives, the package name, and version.
We use a malware package name and a version to confirm its malware sample in the large-scale malware dataset.
Specifically, the large-scale dataset has 5,890 malicious packages, of which 572 only appear in malware analysis reports.
The reason is straightforward: not all malware packages were analyzed by security professionals. Only 9.7\% malware packages were analyzed in the security reports.

\textbf{Dataset Correctness}. 
We assume that online sources are reliable and trustworthy and have a good reputation in the security community.
Some packages may be false positives when a legitimate package is falsely reported as malware. 
It is impractical to manually check each malware package to remove false positives from the dataset.
For malware analysis reports, few works provide security analysis reports about SSC attacks and OSS malicious packages.
We release our data sets through the private GitHub repository (application access), where we receive community feedback and filter out false positives.

\textbf{Dataset Transparency}.
To guarantee the reproducibility of OSS malware, we provide details on the transparency of our datasets. 
Due to page limits, we present only a summary of our datasets in the paper; the detailed datasets are available in a private GitHub repository.
We build a private GitHub repository to publish all malware package names (sources) with their signatures (e.g., MD5 hashes).

\textbf{Malware Diversity}.
A large number of malware packages do not imply a high diversity of malware packages, where many malware packages may overlap each other. 
Many prior works~\cite{bailey2007automated, bayer2009scalable, jang2011bitshred} leveraged malware families to organize malware samples, where the malware samples have similar behaviors and attack techniques in the same malware family. 
However, most OSS malware belongs to interpreted malware, which differs from compiled malware. 
OSS malware uses an SBOM to organize the components; no binary malware samples exist.
The OSS malware family is unavailable because the interpreted malware does not have binary-level polymorphism.
Thus, we use malware signatures to remove duplicated malware packages.

\section{Evaluation}

In this section, based on the dataset with the ground truth (Table~\ref{tab:report:source}), we conduct real-world experiments to evaluate the effectiveness of {\tech}.
Our evaluation aims to address the following research questions. 

\noindent \textbf{• RQ1 (Performance)}: How accurately can {\tech} identify TTPs? 

\noindent \textbf{• RQ2 (Model Versatility)}: How does {\tech} perform when applied on different LLMs?

\noindent \textbf{• RQ3 (Comparison)}: How does the performance of {\tech} compare to existing tools? 

\noindent \textbf{• RQ4 (Missing Rate)}: Is there a possibility for {\tech} to miss real TTPs?

\subsection{Implementation}

The security report collection is based on an LLM-agent system.
We leverage the LangChain framework~\cite{langchain}, an LLM-powered platform, to implement the web crawler and security report identification.
The LLM agent uses a scrapy framework-based tool~\cite{scrapy} API to collect webpages and a beautifulSoup-based tool~\cite{beautifulsoup} API to extract webpage content. 
We leverage OpenAI API to access the GPT-4 model to identify whether a webpage is an analysis report about OSS malware.
Our LLM-agent system obtains 1,366 security analysis reports.

The malware collection is a manual process, and we leverage the report dataset to collect malware packages. 
We manually extract malware package names and versions from the security analysis reports.
The malware dataset covers 572 malware packages.

We have implemented a prototype system for {\tech} based on the LangChain framework~\cite{langchain}. 
Malware content extraction is based on an LLM-based agent.
We provide LLMs with a prompt (encapsulating instruction, the context, and the requirements) for each subtask.

\subsection{Performance (RQ1)}

We cannot directly compare {\tech}'s results with the ground truth for two reasons. First, security reports' TTPs are written by security professionals in the natural language.
Second, generated TTPs contain detailed and specific information about the attack vectors, while the ground truth only contains the attack vector names.  
Therefore, we propose two metrics to evaluate the precision of {\tech}: coverage rate ($CR$) and sequence accuracy ($SA$).
The $CR$ is defined as:
$$CR = \frac{Set_{g}\cap Set_{r} }{Set_{r}} $$
where $Set_{g}$ is the set of attack vectors generated by {\tech}, and $Set_{r}$ is the set of attack vectors in the ground truth.
$CR$ represents the accuracy of {\tech} in extracting common attack vectors from malware packages.
The $SA$ is defined as:
$$SA = \frac{LCS(Set_{g}, Set_{r}) }{Set_{r}}$$
where $LCS(\cdot)$ is the longest common subsequence of two sequences.  
A subsequence is any subset of the elements of a sequence that maintains the
same relative order.
For example, if $Set_g$ is \{typosquatting $\rightarrow$ imitated-url $\rightarrow$ fake-contact $\rightarrow$ pre-install $\rightarrow$ install $\rightarrow$ cmd $\rightarrow$ evasion $\rightarrow$ data\} and $Set_r$ is \{install $\rightarrow$ evasion $\rightarrow$ conceal $\rightarrow$ data\}, the $LCS(Set_{g}, Set_{r})$ is \{install $\rightarrow$ evasion $\rightarrow$ data\}.
$LCS(\cdot)$ represents how accurately {\tech} extracts the execution order from a malware package.

Table~\ref{tab:precision} shows the performance of {\tech} on the dataset, covering 54 reports and 365 PyPI malware packages.
We observe that the average $CR$ is 0.90 and the $SA$ is 0.99.
The $CR$ is high, indicating that {\tech} can accurately extract common attack vectors from a malware package.
The $SA$ is high, indicating that {\tech} can accurately obtain the execution order of the attack vectors from a malware package.

\begin{table}[!t] \small
    \centering
    \caption{The performance of {\tech} to generate the malware TTP.}
    \label{tab:precision}
    \begin{tabular}{ c c c c c  }
    \hline
    Site                & \makecell{Report\\Num.}  & \makecell{Pkg\\Num.}  & $CR$           & $SA$ \\\hline
    blog.phylum.io      & 11            & 95            &    0.93       &    1.00    \\
    blog.sonatype.com   &  11           & 87           &   0.92        &    0.99   \\
    checkmarx.com/blog  &  9           & 35             &   0.71        &     1.00     \\
    lwn.net             &  1           & 5             &    0.94       &      0.99     \\
    medium.com          &  11          & 84            &   0.85        &     1.00   \\
    thehackernews.com   &  8           & 55            &   0.82        &      0.97     \\
    theregister.com     &  1           & 2             &    1.00          &      1.00    \\
    bertusk.medium.com  &  2           & 2             &   1.00        &     1.00 \\\hline
    Total               & 54            & 365           & 0.90       & 0.99    \\
    \hline
    \end{tabular}%
\end{table}

\begin{table}[!t] \small
    \centering
    \caption{The precision of {\tech} among different LLMs.}
    \label{tab:llm}
    \begin{tabular}{ c c c c }
    \hline
                     & $CR$ & $SA$ \\\hline
        LLaMA2~\cite{llama}       &  0.69         &     0.67         \\
          QWen~\cite{tongyi}       &  0.73         &      0.85       \\
        Gemini pro~\cite{gemini}   &  0.75         &     0.85        \\
        GPT-3.5~\cite{openai}      &  0.71         &     0.73        \\
        GPT-4.0~\cite{openai}      &  0.90         &    0.99      \\\hline

    \end{tabular}%
\end{table}

\subsection{Model Versatility (RQ2)}

We evaluate {\tech} on different LLMs to test their versatility.
We select five LLMs for the performance evaluation of {\tech}.
(1) LLaMA2~\cite{llama} is an open source of LLM, with 2 trillion training tokens.
(2) QWen~\cite{tongyi} is an LLM with 72 billion parameters developed by the Alibaba Group.
(3) Gemini Pro~\cite{gemini}  is an LLM from Google that supports a context window of 32k tokens.
(4) GPT-3.5~\cite{openai} is the LLM from OpenAI, and we test with version gpt-3.5-turbo-0613.
(5) GPT-4.0~\cite{openai} is the latest version of LLM from OpenAI, and we test version gpt4-0613.
All LLMs provide APIs to send a prompt and receive a response.

Table~\ref{tab:llm} shows the performance of {\tech} using different LLMs. 
The average $CR$ is 0.76 and the $SA$ is 0.82. In addition, we have the following findings. 
First, the performance of {\tech} is not stable over different LLMs. GPT-4.0 is the most advanced and possibly the most powerful LLM, achieving the best performance, nearly 0.90 $CR$ and 0.99 $SA$.
Other LLMs have similar performance, far worse than GPT-4.0. 
Second, the non-deterministic nature of LLMs brings inconsistencies and non-reliable results.
Third, the performance of {\tech} highly relies on the LLM selection. 
All LLMs (LLaMA2, QWen, Gemini pro, and GPT-4.0) are black-box models, and we cannot control their responses.
In our controlled experiments, we leverage LLaMAfile to train a local LLM whose performance is not acceptable, nearly 0.4 $CR$. 
Similar to many prior works~\cite{feng2023prompting, li2023hitchhiker, pearce2023examining, ullah2023can}, leveraging the most mature and emerging LLM is a better choice than using fine-tuned local LLM to solve a specific security task.

\subsection{Comparison (RQ3)}

\begin{table}[!t] \small
    \centering
    \caption{The performance of {\tech} on existing tools.}
    \label{tab:tool}
    \begin{tabular}{c c c }
    \hline
       Tool        &   Detection Approach      &    $CR$             \\ \hline

    GuardDog~\cite{guarddog}        &   Offline\&Rule       &   0.98       \\ 
    Semgrep~\cite{semgrep}      &   Offline\&Rule      &    0.95           \\  
    Aura~\cite{aura}   &   Online\&Rule     &    0.83             \\ 
    Snyk Code Test~\cite{snyk-online}  &   Online     &   0.93         \\ 
    \hline
        Average      &                   &  0.92          \\  \hline
    \end{tabular}%
     \vspace{-4MM}
\end{table}

Further, we select four existing tools to compare the performance of {\tech}, and the results are presented in Table~\ref{tab:tool}.  
All tools leverage a set of heuristics on the source code and the package metadata to detect malicious behaviors of packages. 
(1) Semgrep~\cite{semgrep} uses heuristic rules to scan source code to identify supply chain security issues. 
(2) GuardDog~\cite{guarddog} is a tool for identifying malicious PyPI and NPM packages.
GuardDog reuses Semgrep heuristic rules and supplements its rules on the package metadata.
(3) Synk Code Test~\cite{snyk-online} is an online test platform to fix vulnerabilities in source code, dependencies, and container images.
Synk Code Test is a commercial and proprietary tool, and we do not know how it works or what heuristic rules are.
(4) Aura~\cite{aura} is a static analysis framework to detect the threat of malicious packages and vulnerable code published on the PyPI ecosystem.
Those tools support many formats, such as pretty plain text, JSON, and SQLite.
Table~\ref{tab:tool} shows the performance of {\tech} on the existing tools.
We assume that those tools' results are the ground truth and calculate the $CR$ of {\tech} on the ground truth.
The average $CR$ is 0.92, indicating that {\tech} extracts common attack vectors from the malware package with high accuracy.

We cannot compare the $SA$ of {\tech} on tools because their outputs are security alerts without execution sequences. 
Those tools first extract static indicators of compromise (IOC) or signatures from malware packages, then feed the IOCs into their heuristic rules to aid possible security alerts in OSS packages.
However, security alerts are insufficient to understand malware's capabilities, and heuristic rules do not yet exhibit the logic behind the malware. 
In addition, a rule is written by security professionals, which is not scalable to numerous malware packages.
To the best of our knowledge, no existing tools are designed to provide fully automated analysis for TTPs of malware packages without any manual efforts.

\begin{figure}
    \centering
    \includegraphics[width=2.0in ]{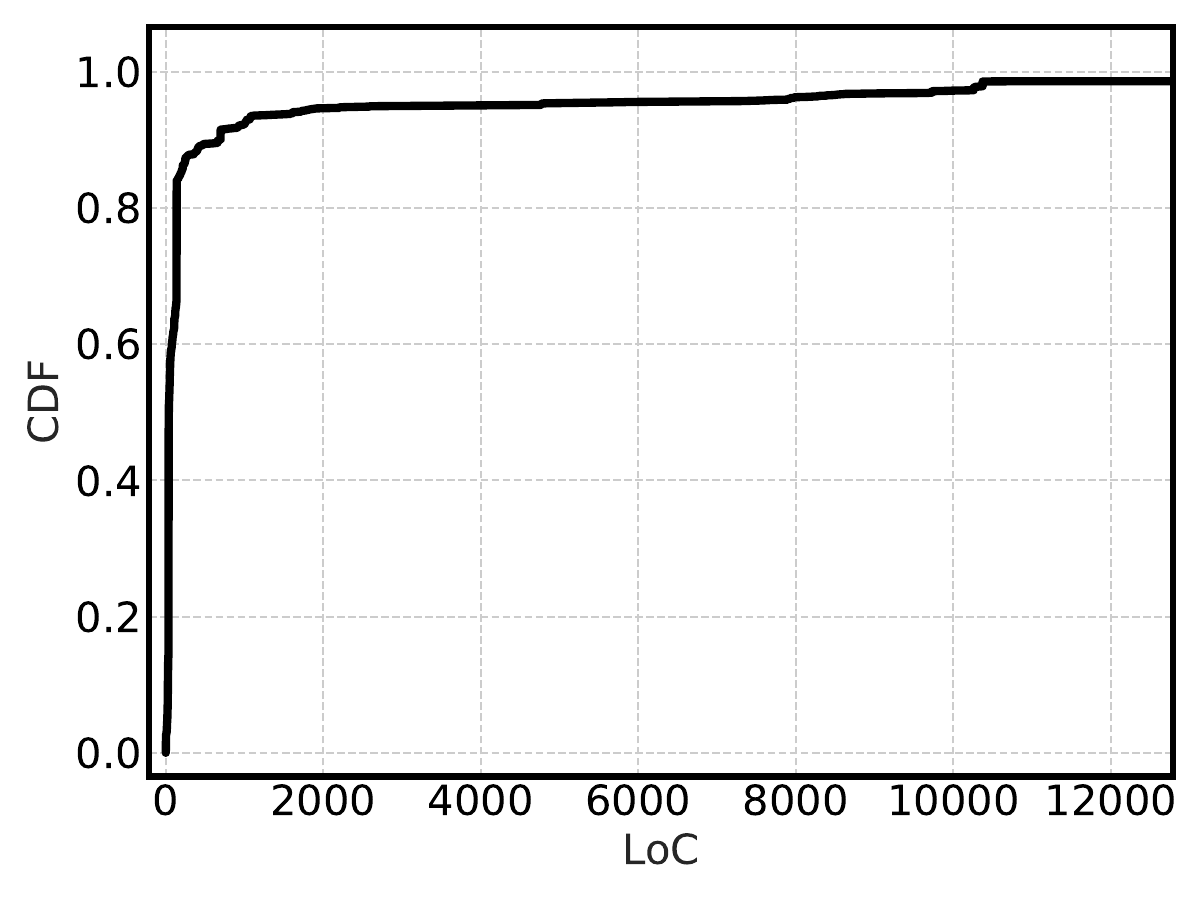}
    \caption{The distribution of malware code's LoC.}
    \label{fig:loc}
     \vspace{-4MM}
\end{figure}

\subsection{Missing TTP (RQ4)}
Table~\ref{tab:report:source} does not present false negatives about malware packages, as we cannot obtain false negatives caused by {\tech}. 
We use 3,700+ PyPI malware packages in the wild (Table~\ref{tab:dataset:malware}) to evaluate the missing TTPs. After manually checking those PyPI malware packages, we found three cases that cause the missing TTP of {\tech}: the lack of package metadata, oversimplified malware code, and complicated malware code.

The first case is the lack of package metadata.
As mentioned above, the package metadata is the key to identify the deceptive tactic TTPs of malware packages. 
The plausible cause is that malware developers are aware of the importance of package metadata, and they intend to leverage the package metadata to deceive the benign developers or users.
We observe that 32 malware packages lack the package metadata, less than 1\% (32/3,735). 
Table~\ref{tab:missing:1} in Appendix lists 32 malware packages without package metadata, and {\tech} cannot identify their deceptive TTPs.

A malware package is converted into its metadata and source code. 
We use the lines of code (LoC) to represent the complexity of the malware code.
Figure~\ref{fig:loc} depicts the overall CDF of LoC of 3,700+ PyPI malware.
There are 3,517 (94.2\%) malware packages with less than 2,000 LoC and 3,323 (89\%) malware packages with less than 500 LoC.
Today's LLM prompt has enough large window sizes to encapsulate malware code with 2,000 LoC.
In this case, {\tech} can identify the execution tactics of interpreted malware.

The second case is that malware uses oversimplified code.
When a malware code contains very few code snippets, {\tech} may not be able to identify execution tactics. 
We observe that 65 (1.74\%) of the malware packages contain zero LoC in their code, and 32 (0.86\%) of the malware packages contain one LoC.
Table~\ref{tab:missing:3} in the Appendix lists those 65 malware packages with empty code.
The empty code is the cause of the missing TTP, leading to {\tech}'s limitation.
Those malware packages belong to a pure metadata package without embedding any malicious code or any malicious behavior.

The third case is that malware uses complicated code.
When a malware package contains too long code snippets, {\tech} may not be able to identify execution tactics. 
We observe that 54 (1.44\%) of the malware packages contain more than 15,000 LoC, and 7 of the malware packages contain more than 20,000 LoC. 
Table~\ref{tab:missing:2} in Appendix lists 54 malware packages over 15,000 LoC.
Those packages inject malicious code snippets into existing open-source packages or just copy the entire code from existing open-source packages.
Hiding malicious code in a large code base may be an evasion approach to bypassing security analysis.
For instance, malware `awscl-1.27.67' has 38,320 LoC and malware `pythonkafka-1.3.5' has 32,128 LoC.
The complicated code is one of the causes of the missing TTP, leading to {\tech}'s limitation.

Overall, {\tech} leverages the metadata and malware code to extract deceptive and execution TTPs.
The lack of metadata, too simple code, and too long code are the main causes of the missing TTP of {\tech}.

\begin{figure}[!t]
    \centering
    \includegraphics[width=3.3in ]{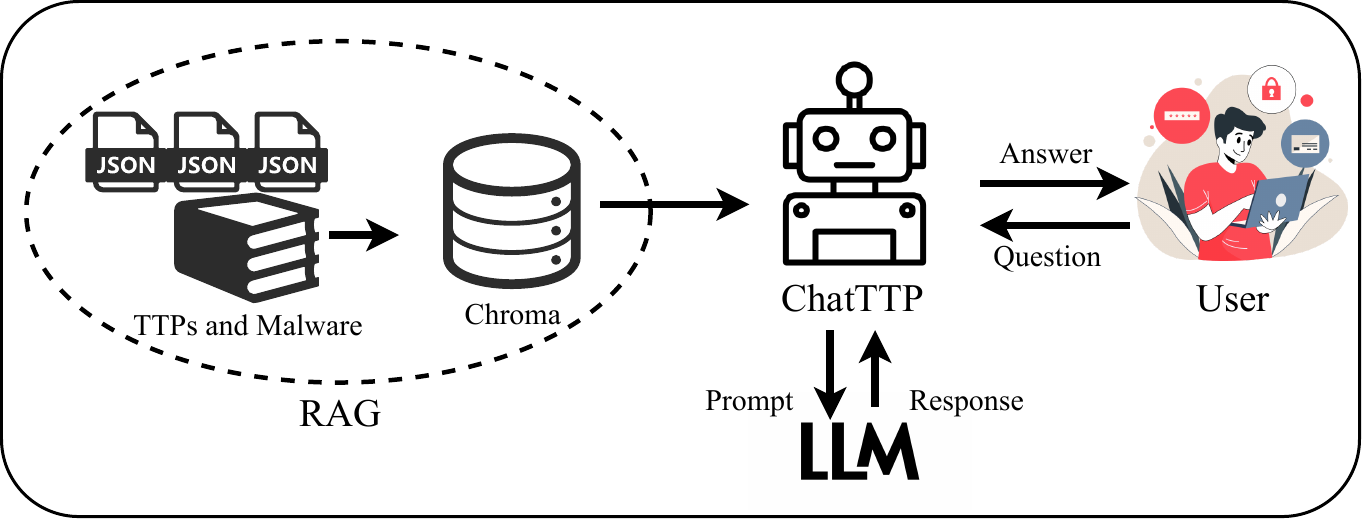}
    \caption{The architecture of the ChatTTP.}
    \label{fig:chatttp}
    \vspace{-3MM}
\end{figure} 

\section{Insights into TTP}

In this section, we use a large dataset (Table~\ref{tab:dataset:malware}) to generate 3,735 TTPs for PyPI malware packages.
First, we implemented an LLM Chatbot (called ChatTTP) based on 3,700+ malware TTPs.
Then, we conducted a quantitative analysis of the malware TTPs in the wild.

\subsection{ChatTTP: Malware Analysis QA}  

ChatTTP is a Chatbot based on LLMs to demonstrate the advantages of {\tech}. 
Specifically, we leverage several frameworks to implement ChatTTP, including LangChain~\cite{langchain}, OpenAI's GPT-4.0, and the Chroma vector database. 
Figure~\ref{fig:chatttp} depicts the architecture of  ChatTTP that helps users retrieve and analyze malware package information.
As a critical component of ChatTTP, the retrieval augmented generation (RAG) provides external knowledge to LLMs.
Given a QA prompt, RAG retrieves the most relevant documents or passages from the dataset.
More specifically, RAG follows two steps: (1) indexing documents and (2) retrieving the relevant document.

\textit{Document}. 
Each malware package is stored as a document in the JSON format. 
Each document contains a package name, a version, an ecosystem name, deceptive and execution TTPs, and analysis results.
Specifically, 3,700+ documents (Table~\ref{tab:dataset:malware}) are stored in the Chroma vector database as external knowledge.
It is unnecessary to split the document into chunks, because it is short and the document size fits the full post in the token size of LLMs.
Then, we leverage OpenAI's GPT-4.0 to convert the content of each document into an embedding vector. 
The Chroma vector database stores and builds the index of all documents and their embedding vectors.

\textit{Retriever}. 
Given a QA prompt, RAG uses a retriever to find information about malicious packages in the Chroma vector database.
First, we compute the prompt's embedding vector via the OpenAI's GPT-4.0 model.
To find the most relevant documents, the similarity between the prompt vector and the vectors from the Chroma vector database is calculated.
Specifically, we reuse Facebook AI Similarity Search (FAISS)~\cite{faiss} to perform the similarity calculation and search. 
The FAISS advantage is that the similarity between two vectors follows a Gaussian distribution in a high dimensional space, and we can use the clustering technique to improve the efficiency of similarity search.

\begin{figure}[!t]
    \centering
    \includegraphics[width=2.0in ]{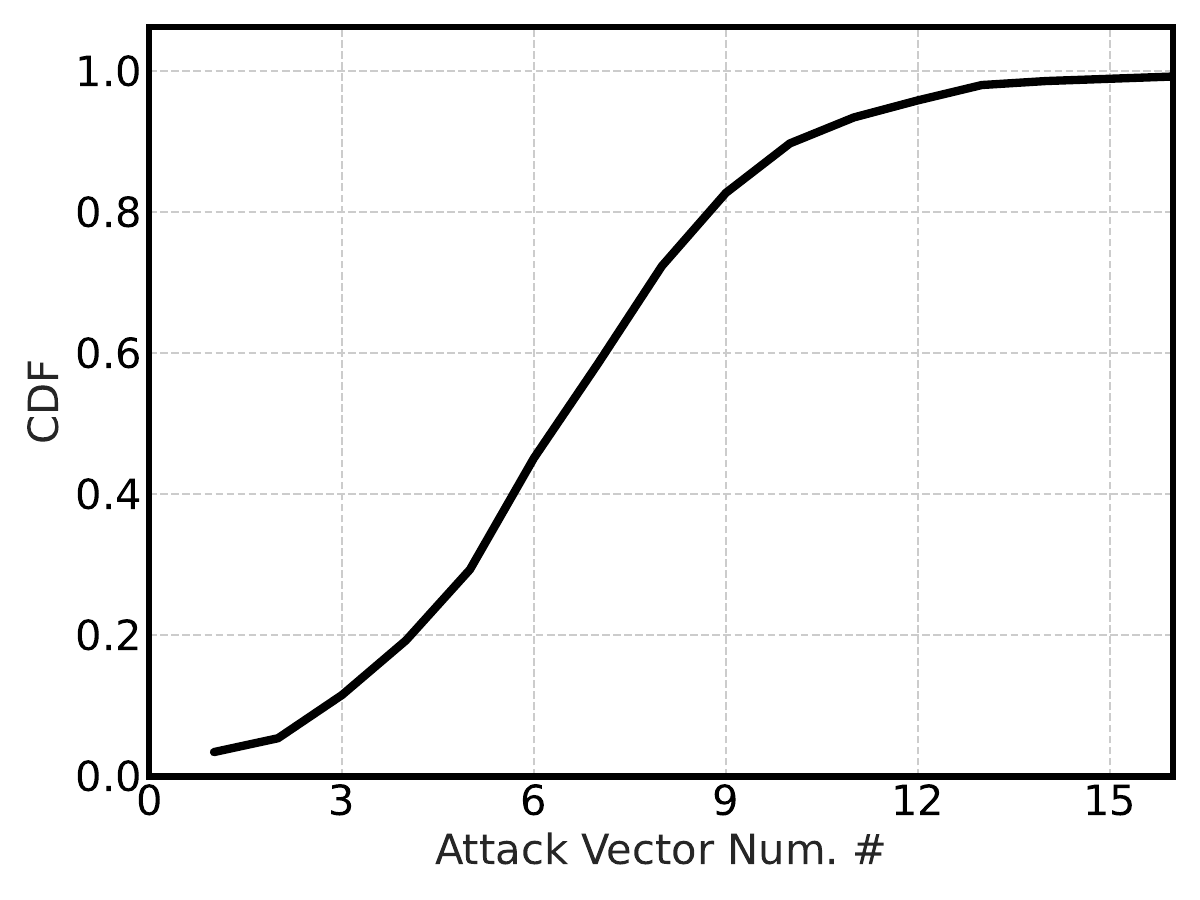}
    \caption{The distribution of attack vector number for 3,700+ PyPI malware's TTPs.}
    \label{fig:ttp:cdf}
    \vspace{-3MM}
\end{figure} 

\textbf{Usage}.
ChatTTP can answer questions about malicious packages, and we release it as a application demo in the website\footnote{Chatbot Demo in anonymous submissions: \url{https://genttp.streamlit.app/}}.
The QA prompt template is shown in Table~\ref{tab:chatbot} (Appendix). 
When a user asks ChatTTP a question, the retriever component of the vector database retrieves relevant documents. 
LLMs use relevant documents (malware's TTP knowledge) to generate responses.

Overall, TTPs of malicious packages bring three advantages to the security community.
First, malware analysis is critical for staying ahead of adversaries, and today's adversaries typically develop an arsenal of attack vectors rather than a single attack vector that they routinely use to launch their attacks. 
Recognizing and tracking a malware's strategy can help organizations better defend against existing or upcoming targeted attacks. 
Second, malware TTPs provide a practical context for characterizing an adversary's objective and malicious behaviors during running/installing a malware package, such as code obfuscation, downloading files, executing files, and exfiltrating data. 
Third, TTPs can be used as a reference to a malware analysis report. 
A Chatbot based on numerous malware's TTPs can provide a deep understanding of malware's capabilities far beyond the malware detection of existing tools (e.g., OSSGadget~\cite{gadget} or Semgrep~\cite{semgrep}).

\subsection{TTP in the wild}

{\tech} extracts all malware TTPs in the PyPI ecosystem and provides a comprehensive view of the TTP landscape in the wild.
Figure~\ref{fig:ttp:cdf} depicts the CDF of TTP length, where the X-axis is the number of attack vectors per TTP.
We can observe that 82.7\% of TTP length is less than 9.0, and 89.7\% of TTP length is less than 10.0.
This indicates that most malware packages contain less than 10 attack vectors.
The maximum number of attack vectors in a malware package is 18.
We observe that our extracted TTPs have more common and detailed attack vectors than TTPs from security analysis reports, and criminals have an arsenal of attack vectors to develop malicious packages.

\begin{table*}[!t]\small
\vspace{-0.1in}
    \centering
    \caption{The top 10 TTP among 3700+ malicious packages. \textbf{TS} represents typosquatting, \textbf{IV} represents imitated version, \textbf{FD} represents fake description, \textbf{FC} represents fake contact, \textbf{IU} represents imitated URL, \textbf{MD} represents malicious dependence, \textbf{Pre} represents preinstall, \textbf{CMD} represents cmd, \textbf{EVA} represents evasion, and \textbf{Con} represents conceal, \textbf{Remote} represents URL/IP/port.}
    \label{tab:top:ttp}
    \begin{tabular}{ l c r  }
    \hline
       TTP              &  Malware Num. \# &  Attacker Intent \\\hline
    \{TS$\rightarrow$IV$\rightarrow$FD $\rightarrow$FC$\rightarrow$Pre$\rightarrow$CMD$\rightarrow$EVA$\rightarrow$Con\} &	716 &	\makecell[r]{Trojan, Spyware, Downloader, and Backdoor}\\

    \{TS$\rightarrow$IV$\rightarrow$ FD$\rightarrow$CMD$\rightarrow$EVA$\rightarrow$Con\} &	455 & \makecell[r]{Trojan type and Backdoor}\\

    \{TS$\rightarrow$IV$\rightarrow$FD$\rightarrow$Pre$\rightarrow$CMD$\rightarrow$EVA$\rightarrow$Con\} &	351 & \makecell[r]{Downloader} \\

    \{TS$\rightarrow$IV$\rightarrow$FD$\rightarrow$MD$\rightarrow$IU$\rightarrow$Pre$\rightarrow$CMD$\rightarrow$EVA$\rightarrow$Con\} &	274 & \makecell[r]{Trojan, Spyware, Backdoor and   Downloader}\\

    \{TS$\rightarrow$EVA$\rightarrow$Con$\rightarrow$CMD\} &	225 & Trojan type, backdoor \\

    \{TS$\rightarrow$IV$\rightarrow$FD$\rightarrow$FC$\rightarrow$CMD$\rightarrow$EVA$\rightarrow$Con\} &	163 & Trojan type\\

    \{TS$\rightarrow$IV$\rightarrow$ FD$\rightarrow$MD$\rightarrow$IU$\rightarrow$FC$\rightarrow$Pre$\rightarrow$CMD$\rightarrow$EVA$\rightarrow$Con\} &	137 &  \makecell[r]{Trojan, Spyware, Backdoor,  and Downloader}\\

    \{TS$\rightarrow$IV$\rightarrow$FD$\rightarrow$FC$\rightarrow$Pre$\rightarrow$CMD$\rightarrow$EVA$\rightarrow$Con$\rightarrow$Remote\} &	102 & \makecell[r]{Trojan, Spyware, Backdoor,  and Downloader}\\

    \{IU$\rightarrow$FC$\rightarrow$CMD$\rightarrow$EVA$\rightarrow$Con\} &	63 & Trojan\\

    \{TS$\rightarrow$IV$\rightarrow$FD$\rightarrow$IU$\rightarrow$FC$\rightarrow$Pre$\rightarrow$CMD$\rightarrow$EVA$\rightarrow$Con\} &	49 & Trojan type, Backdoor\\

                     \hline

    \end{tabular}%
\end{table*}

\begin{figure}[!t]
    \centering
    \includegraphics[width=3.3in ]{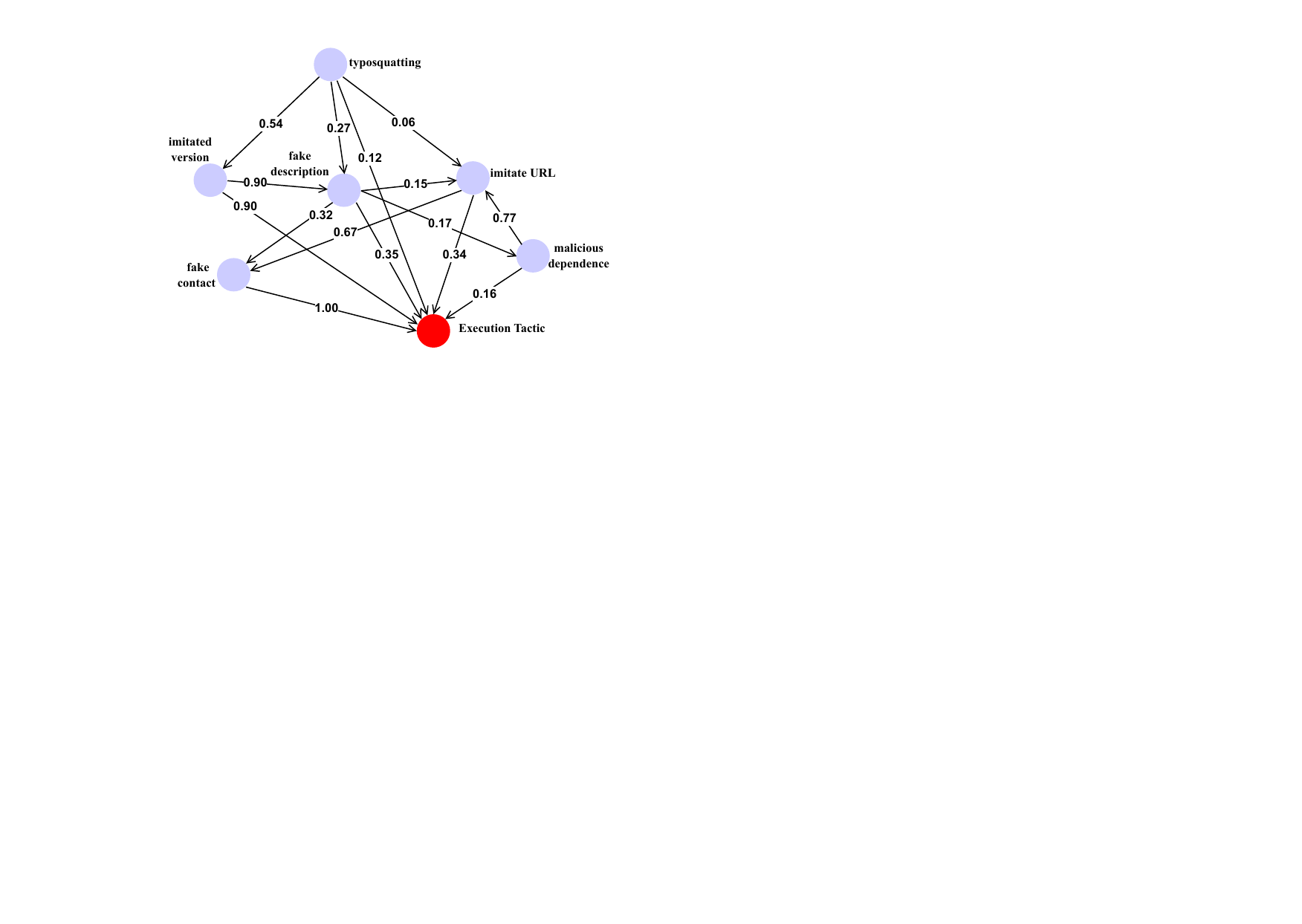}
    \caption{The state transition graph for the deceptive tactic in the metadata.}
    \label{fig:transmittion:dec}
    \vspace{-3MM}
\end{figure}

Each TTP describes how an attacker uses techniques to implement deceptive and malicious activities of the malware.
If two malicious packages share the same TTP, we merge them into one TTP.
Table~\ref{tab:top:ttp} lists the top 10 TTPs with the most malware packages. 
Those top 10 TTPs cover 2,535 (67.9\%, 2,535/3,735) different malicious packages.
The TTP \{TS$\rightarrow$IV$\rightarrow$FD$\rightarrow$FC$\rightarrow$Pre$\rightarrow$CMD$\rightarrow$EVA$\rightarrow$Con\} covers the largest number of malware packages, close to 716.
It uses four deceptive attack vectors and five execution attack vectors to achieve the malware objective.
The TTP \{TS$\rightarrow$IV$\rightarrow$ FD$\rightarrow$CMD$\rightarrow$EVA$\rightarrow$Con\} covers the second-largest number of malware packages, close to 455, which contains three deceptive attack vectors and three execution attack vectors.
Among those TTPs, we observe that TS (typosquatting) and IV (imitated version) are the most common attack vectors for malicious packages. 
The reason is that TS/IV is the first step in a TTP to deceive users into installing malware. 
Similar to code reuse in software development, malware authors also reuse similar attack vectors to build their malware. 
From the attacker's perspective, attack vector reuse is common in today's malware development for improved efficiency. 
In addition, attack vector reuse may lead to an increase in the number of malware and attack campaigns.

\finding{ 
Many OSS malicious packages share a relatively stable TTP, suggesting that attack vector reuse is common in today's malware development. 
A deceptive TTP is common in OSS malware. 
}

\begin{figure}[!t]
    \centering
    \includegraphics[width=3.2in ]{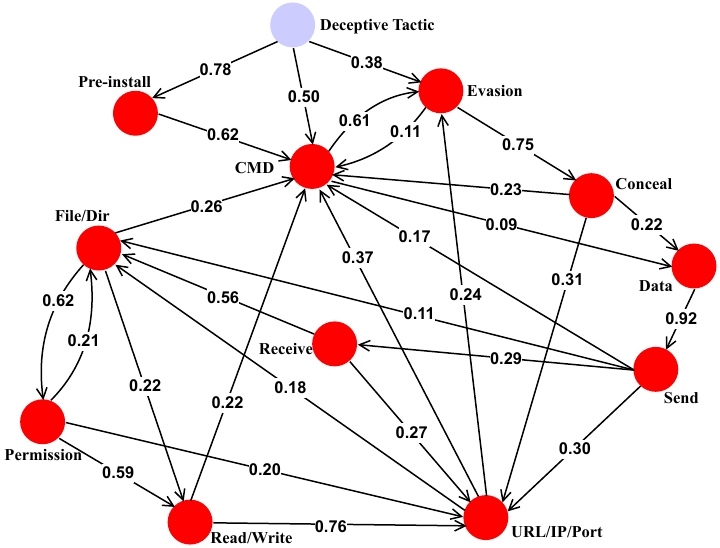}
    \caption{The state transition graph for the execution tactic in the malware code.}
    \label{fig:transmittion:exec}
    \vspace{-3MM}
\end{figure} 

Based on 3,700+ TTPs, we construct a state transition graph ($G$), which consists of two sets: a set of nodes ($V$) and a set of edges ($E$).
A node in $G$ represents an attack vector and an edge in $G$ represents the state transition between two attack vectors. 
An edge weight is the transition probability between two attack vectors, denoted as:
$$w(AV_j|AV_i) = \frac{|AV_i \rightarrow AV_j| }{|\sum_{k = 1}^{n} AV_i \rightarrow AV_k|}, $$
where the $AV_i$ indicates an attack vector, and the $AV_i \rightarrow AV_j$ is the state transition from $AV_i$ to $AV_j$ in a TTP, and $n$ is the total number of attack vectors. 
Formally, the state transition graph can be represented as $G$ = ($V$, $E$, $W$). 
Note that $G$ is a weighted directed graph, and circles are allowed. 
In short, the state transition graph $G$ facilitates the analysis on interpreted malware.

Figure~\ref{fig:transmittion:dec} shows the state transition graph of the deceptive TTPs, and Figure~\ref{fig:transmittion:exec} depicts the state graph of the execution TTPs.
Each arrow $\rightarrow$ represents the state transition from one attack vector to another, and the number on the arrow represents the transition probability. 
For example, two states are in conjunction with initial access to download code and lateral movement to execute code in the local system.
Specifically, those attack vectors are used in conjunction with enabling the malware's ability.
We observe several high probabilities in the deceptive tactic, including 0.67 (imitated URL$\rightarrow$ fake contact), 0.90 (imitated version$\rightarrow$fake description), and 0.77 (malicious dependency$\rightarrow$imitated URL).
The execution tactic has several combinations with high probability, such as 0.92 (Data$\rightarrow$Send), 0.62 (Pre-install$\rightarrow$CMD), and 0.76 (Read/Write$\rightarrow$url/ip/post).
A high probability indicates that malware authors tend to combine two attack vectors in the same malicious package. 
By contrast, a low probability indicates that two attack vectors rarely co-exist in a malicious package. 
Note that the state transition graph is relatively stable over time, even with the increasing occurrence of malware and attack campaigns.

In a graph $G$, a path is a sequence of edges, each one incident to the next, as the main parameter of interest for a TTP.
Formally, a path length $n$ from node $v_0$ to node $v_n$ is a sequence of $n$ edges \{$e_1$, $\dots$, $e_n$\} in the graph, where $e_i$ = ($v_{i-1}$, $v_i$).
Specifically, a malware TTP is mapped to a path (a node sequence) in the graph, where its probability is equal to the path probability by multiplying the transition probability in edges.

\finding{
A state transition graph depicts how malware developers utilize existing techniques and tactics to achieve malicious objectives.
A TTP reflects the characteristics of a malware-based attack via a path in a state transition graph.}

The industry focuses on malware analysis to understand the malware behaviors and attacker intents behind the malware.
Here, we leverage 3,700+ malicious packages as ground truth to understand the relationship between TTPs and attacker intents.
Similarly, we extract attacker intents behind 3,700+ malicious packages by using corresponding TTPs as follows.
(1) First, we select a TTP and find its corresponding malware packages. 
(2) Second, in terms of package names and versions, we find their security reports.
(3) Third, we manually analyze the security reports to confirm the mapping between the TTP and the attacker's intent.
Table~\ref{tab:top:ttp} lists the top 10 TTPs for revealing the attacker intents and the malware patterns.
We observe that a TTP contains multiple intents in a malicious package.
The first category of malware intent (Spyware) is to steal sensitive information, such as user credentials, financial information, and personal information.
It is consistent with the practical observation that today's OSS malicious packages act as trojan-type malware in victims' systems. 
The second category (Backdoor) is to use multiple evasion strategies to avoid detection by security software.
The third category (Downloader) involves using IP/URL to download and execute malicious code from the Internet.

\finding{
An attacker's intent behind the malware is linked to a TTP, which is important for defense and security analysis.}
\section{Discussion and Limitation}

\textbf{Threats to Validity}. 
{\tech} suffers from three threats to validity. 
First, our ground-truth dataset relies on security analysis reports written by security professionals. Although we manually checked the reports, there is still a possibility that they contain false positives. 
Second, our large-scale malware dataset lacks the ground truth labels for TTPs. Our insights on malware TTPs may be affected by the false positives and false negatives of  {\tech}.
We rely on the LLM to extract the TTPs, and the LLM is a black box.
Third, it may raise concerns about the unbiasedness and stability of our analysis results.
New malicious packages and new attack vectors threaten the validity of malware TTPs. 
So far, 3700+ PyPI malware packages have unique signatures (MD5 hash value), and many malware packages are similar to each other.
We do not filter out similar packages because there is a lack of similarity in the interpreted malware.

\textbf{Malware Package Limitation}. 
Our approach leverages the package metadata to extract deceiving tactics and uses malware code to extract execution tactics.
If the package metadata or malware code is missing, our {\tech} cannot extract the TTPs.
There are many malware packages that cannot be converted into package metadata and malware code.
We can only obtain the binary file of malware, e.g., windows PE format. In this case, {\tech} cannot analyze the binary of the malware.
In addition, malware will continue to evolve and malware developers will use more complicated evasion techniques. 
The security community should update their knowledge on advanced ways to analyze and defeat malware.

\textbf{Extension to More Ecosystems}.
So far, our {\tech} only supports extracting TTPs from PyPI malware packages. 
The reason is that we use a Python code interpreter, not to adapt to other ecosystems. 
Our approach is general to any interpreter with extending engineering efforts. 
In our future work, we would extend our approach to more ecosystems, e.g., NPM, RubyGems, and Maven. 
We have the largest number of malware packages from the PyPI ecosystem.
Another issue is that malware packages are missing when they come from different ecosystems. 
A reliable and high-quality malware dataset is vital for analyzing malware packages, where the security community should work together to collect missing and unknown malware packages in the wild.

\section{Related Work}

In this section, we survey related work in three aspects: LLM, malware analysis, and the OSS ecosystem.

\textbf{LLM} is primarily based on the Transformer architecture with extensive pre-training on large-scale training data, and it has demonstrated remarkable advances across various domains. 
Several prior works~\cite{wei2022chain, yao2023tree, yao2022react} proposed a planner to extend LLMs' capabilities for dealing with complex tasks, such as generalization and reasoning skills. 
Task decomposition techniques divide a complicated task into small pieces, such as Chain-of-thought~\cite{wei2022chain} and Tree-of-thought~\cite{yao2023tree}. 
Yao et al.~\cite{yao2022react} proposed a general strategy of self-reflection (called ReAct) based on the feedback of LLMs to improve their reasoning skills.
Similarly, Reflexion~\cite{shinn2023reflexion} and Chain of Hindsight~\cite{liu2023languages} use human feedback (e.g., errors) to fine-tune LLMs.
Another distinct approach is to use external resources to build an autonomous agent, enabling LLMs to interact with the environment (tools or APIs).
Weng~\cite{llmagent} summarized that an LLM-based agent needs task decomposition and self-reflection.
Wang et al.~\cite{wang2023voyager} proposed an embodied lifelong learning agent based on LLMs.

Meanwhile, LLMs can be applied in the software engineering domain, such as OpenAI Codex and GitHub Copilot.
Many prior works leveraged LLMs to resolve specific tasks in the software engineering domain~\cite{ma2023scope, sun2023automatic, pei2023can, xia2023keep, chen2021evaluating, fan2023large}.
Ma et al.~\cite{ma2023scope} and Sun et al.~\cite{sun2023automatic} explored the capabilities of LLMs when performing various program analysis tasks, including control flow graph construction, call graph analysis, and code summarization. 
Pei et al.~\cite{pei2023can} used LLMs to reason about loop invariants and automated repair of programs.
There are many prior works using LLMs to resolve specific tasks in the security domain~\cite{pearce2023examining, chen2023teaching, feng2023prompting, ullah2023can, li2023hitchhiker, pearce2022pop}.
Pearce et al.~\cite{pearce2022pop} proposed to leverage LLM to help security professionals reverse engineer the binary application for automatically repairing vulnerabilities.
Li et al.~\cite{li2023hitchhiker} presented LLM capabilities in the static analysis of finding vulnerabilities. 
Feng and Chen~\cite{feng2023prompting} proposed to use LLM to replay Android bugs automatedly. Pearce et al.~\cite{pearce2023examining} examined zero-shot vulnerability repair using LLMs, and Ullah et al.~\cite{ullah2023can} reasoned security vulnerabilities based on LLMs. 
In real-world examples, the LLMs were not as effective as expected for specific security tasks.

\textbf{Malware Analysis} is to gain a deep understanding of malware behaviors. 
Prior research on conventional malware~\cite{bailey2007automated, bayer2009scalable, jang2011bitshred} leveraged the malware samples with corresponding traces to investigate the malware behaviors. 
Bayer et al.~\cite{bayer2009scalable} and Jang et al.~\cite{jang2011bitshred} proposed clustering techniques to partition malware samples into different groups based on their behaviors, and they analyzed the behavioral profiles of malware samples.
Note that we do not use the malware family to group those interpreted malware. 
The reason is that OSS malware belongs to the interpreted malware rather than compiled malware.
The malware family is used to represent similar behaviors of binary-level polymorphism based on the same code base.
VX underground~\cite{vx_underground} is the largest open database of conventional malware, which contains a large collection of malware samples and corresponding code bases. 
Rokon et al.~\cite{rokon2020sourcefinder} proposed SourceFinder to search the GitHub repository based on malware samples and provided a dataset of conventional malware code.
Those databases are designed for compiled malware, but they are not suitable for OSS malware.

In recent years, the research community has paid attention to OSS malware analysis, proposed malware detection mechanisms~\cite{SejiaAdria2022Machinelearn, zhang2020cyber, qian2022malicious, ferreira2021containing, vu2023bad}, analyzed malware behaviors~\cite{wy2022InstalltimeAtt, lidisa2022javabytecode, sejfia2022practical}, and collected malware datasets~\cite{backstabbers-online, Ohmssc2020, guo2023empirical}.
Sejfia et al.~\cite{SejiaAdria2022Machinelearn} presented a machine-learning-based approach for automatically detecting malicious packages.
Zhang et al.~\cite{zhang2020cyber} proposed deep neural networks to analyze the code bases and detect malware packages in the GitHub repository. 
Qian et al.~\cite{qian2022malicious} proposed adversarial contrastive learning to detect malicious repositories in GitHub.
The security policy on OSS registries has also been studied extensively.
Ferreira et al.~\cite{ferreira2021containing} claimed OSS registries could mitigate the impact of malicious packages by taking down those packages.
Wyss et al.\cite{wy2022InstalltimeAtt} tracked SSC attacks in dependencies when installing an NPM package.
Cao and Dolan-Gavitt~\cite{cao2022fork} leveraged the artifacts to find 26 fork repositories among 68 popular cryptocurrency repositories.
Several OSS malware datasets have been available to the public~\cite{backstabbers-online, Ohmssc2020, guo2023empirical}, but their dataset size is much smaller than ours.

\textbf{OSS Ecosystem} is a complex network of software packages and developers, and it is vulnerable to numerous software supply chain attacks. 
Zimmermann et al.~\cite{zimmermann2019small} revealed that the package dependency has become more complicated as time passes, which may be exploited as a launching pad by attackers, thus increasing the package security risk.  
Dann et al.~\cite{dann2021identifying} identified dozens or even hundreds of the direct and transitive dependencies.
Duan et al.~\cite{duan2020towards} proposed a comparative framework for characterizing security features of packages in three OSS ecosystems.
Typosquatting is the most popular attack vector in OSS ecosystems~\cite{vu2021lastpymile, taylor2020spellbound}. 
Ohm et al.\cite{Ohmssc2020} collected, categorized, and manually analyzed a dataset with malicious code from 174 packages that were used for SSC attacks. 
Xiao et al.~\cite{xiao2021abusing} proposed an attack that abuses hidden attributes, which attackers can exploit to obtain confidential data, bypass security checks, and launch denial-of-service attacks. 
Gu et al.\cite{gu2023investigating} built RScouter and found 12 potential attack vectors in the 6 registries, which can be used to distribute malicious packages. 
Ladisa et al.~\cite{ladisa2023sok} presented a systematic literature review in malware analysis, focusing on the techniques and tools used to analyze malware, the datasets used to evaluate these techniques, and the challenges and open issues in the field.

\section{Conclusion}

Interpreted malware plays a critical role in SSC attacks, exploiting an arsenal of attack vectors to achieve malicious purposes.
In this work, we first introduced TTPs to thoroughly characterize the features of various attack vectors in interpreted malware. We then developed {\tech} that leverages the LLM to generate TTPs of malware packages in a zero-shot manner. To validate the efficacy of {\tech}, we collected two datasets, including one dataset with ground truth labels and one large dataset in the wild. 
Our evaluation experiments show that {\tech} can generate TTPs with high accuracy and efficiency.
We have built an LLM-based Chatbot to demonstrate {\tech}'s benefits for malware analysis.
Further, we conduct a quantitative analysis of 3,700+ PyPI malware's TTPs.
Our main findings in the wild are (1) many OSS malicious packages share a relatively stable TTP, even with the increasing emergence of malware and attack campaigns,
(2) a TTP reflects characteristics of a malware-based attack, and (3) an attacker's intent behind the malware is linked to a TTP.


\small
\bibliographystyle{IEEEtranS}
\bibliography{ref}

\begin{thebibliography}{10}
\providecommand{\url}[1]{#1}
\csname url@samestyle\endcsname
\providecommand{\newblock}{\relax}
\providecommand{\bibinfo}[2]{#2}
\providecommand{\BIBentrySTDinterwordspacing}{\spaceskip=0pt\relax}
\providecommand{\BIBentryALTinterwordstretchfactor}{4}
\providecommand{\BIBentryALTinterwordspacing}{\spaceskip=\fontdimen2\font plus
\BIBentryALTinterwordstretchfactor\fontdimen3\font minus \fontdimen4\font\relax}
\providecommand{\BIBforeignlanguage}[2]{{%
\expandafter\ifx\csname l@#1\endcsname\relax
\typeout{** WARNING: IEEEtranS.bst: No hyphenation pattern has been}%
\typeout{** loaded for the language `#1'. Using the pattern for}%
\typeout{** the default language instead.}%
\else
\language=\csname l@#1\endcsname
\fi
#2}}
\providecommand{\BIBdecl}{\relax}
\BIBdecl

\bibitem{aura}
S.~AI., ``{Aura, a static analysis framework.}'' \url{https://github.com/SourceCode-AI/aura}, 2021.

\bibitem{pypi_aliyun}
Alibaba, ``Alibaba cloud pypi mirror for expedited downloads,'' \url{https://mirrors.aliyun.com/pypi/}.

\bibitem{npm_aliyun}
Aliyun, ``Aliyun npm mirror by alibaba cloud,'' \url{https://npm.aliyun.com/}.

\bibitem{bailey2007automated}
M.~Bailey, J.~Oberheide, J.~Andersen, Z.~M. Mao, F.~Jahanian, and J.~Nazario, ``Automated classification and analysis of internet malware,'' in \emph{Recent Advances in Intrusion Detection: 10th International Symposium, RAID 2007, Gold Goast, Australia, September 5-7, 2007. Proceedings 10}.\hskip 1em plus 0.5em minus 0.4em\relax Springer, 2007, pp. 178--197.

\bibitem{bayer2009scalable}
U.~Bayer, P.~M. Comparetti, C.~Hlauschek, C.~Kruegel, and E.~Kirda, ``Scalable, behavior-based malware clustering.'' in \emph{NDSS}, vol.~9, 2009, pp. 8--11.

\bibitem{beautifulsoup}
BeautifulSoup, ``a python package for parsing html and xml documents,'' \url{https://www.crummy.com/software/BeautifulSoup/}, 2012.

\bibitem{ssc_crypt}
Bertus, ``{Cryptocurrency clipboard hijacker discovered in pypi repository},'' \url{https://medium.com@bertusk/cryptocurrency-clipboard-hijacker-discovered-in-pypi-repository-b66b8a534a8}, 2018.

\bibitem{apt}
blackorbird, ``{Interesting APT report, Malware sample and Intelligence Collection.}'' \url{https://github.com/blackorbird/APT_REPORT}, 2024.

\bibitem{cao2022fork}
A.~Cao and B.~Dolan-Gavitt, ``What the fork? finding and analyzing malware in github forks,'' in \emph{Proceedings of the NDSS}, vol.~22, 2022.

\bibitem{chen2021evaluating}
M.~Chen, J.~Tworek, H.~Jun, Q.~Yuan, H.~P. d.~O. Pinto, J.~Kaplan, H.~Edwards, Y.~Burda, N.~Joseph, G.~Brockman \emph{et~al.}, ``Evaluating large language models trained on code,'' \emph{arXiv preprint arXiv:2107.03374}, 2021.

\bibitem{chen2023teaching}
X.~Chen, M.~Lin, N.~Sch{\"a}rli, and D.~Zhou, ``Teaching large language models to self-debug,'' \emph{arXiv preprint arXiv:2304.05128}, 2023.

\bibitem{next-gen}
CISA, ``{Malware Next-GEN system is limited to citizens of the United States. },'' \url{https://www.cisa.gov/resources-tools/services/malware-next-generation-analysis}, 2024.

\bibitem{SSC2017}
T.~Clabur, ``{Typosquatting Attack on npm.}'' \url{https://www.theregister.com/2017/08/02/typosquatting_npm/}, 2017.

\bibitem{npm_cnpmjs}
CNPM, ``Cnpm (china npm) tailored for chinese users,'' \url{https://r.cnpmjs.org/}.

\bibitem{att-ck}
M.~Corporation, ``{MITRE ATT CK is a globally-accessible knowledge base of adversary tactics and techniques based on real-world observations.}'' \url{https://attack.mitre.org/}, 2021.

\bibitem{dann2021identifying}
A.~Dann, H.~Plate, B.~Hermann, S.~E. Ponta, and E.~Bodden, ``Identifying challenges for oss vulnerability scanners-a study \& test suite,'' \emph{IEEE Transactions on Software Engineering}, vol.~48, no.~9, pp. 3613--3625, 2021.

\bibitem{decan2017empirical}
A.~Decan, T.~Mens, and M.~Claes, ``An empirical comparison of dependency issues in oss packaging ecosystems,'' in \emph{2017 IEEE 24th international conference on software analysis, evolution and reengineering (SANER)}.\hskip 1em plus 0.5em minus 0.4em\relax IEEE, 2017, pp. 2--12.

\bibitem{duan2020towards}
R.~Duan, O.~Alrawi, R.~P. Kasturi, R.~Elder, B.~Saltaformaggio, and W.~Lee, ``Towards measuring supply chain attacks on package managers for interpreted languages,'' \emph{arXiv preprint arXiv:2002.01139}, 2020.

\bibitem{fan2023large}
A.~Fan, B.~Gokkaya, M.~Harman, M.~Lyubarskiy, S.~Sengupta, S.~Yoo, and J.~M. Zhang, ``Large language models for software engineering: Survey and open problems,'' \emph{arXiv preprint arXiv:2310.03533}, 2023.

\bibitem{feng2023prompting}
S.~Feng and C.~Chen, ``Prompting is all your need: Automated android bug replay with large language models,'' \emph{arXiv preprint arXiv:2306.01987}, 2023.

\bibitem{ferreira2021containing}
G.~Ferreira, L.~Jia, J.~Sunshine, and C.~K{\"a}stner, ``Containing malicious package updates in npm with a lightweight permission system,'' in \emph{2021 IEEE/ACM 43rd International Conference on Software Engineering (ICSE)}.\hskip 1em plus 0.5em minus 0.4em\relax IEEE, 2021, pp. 1334--1346.

\bibitem{github-advisory}
GitHub, ``{Github Security Advisory Database. },'' \url{https://github.com/advisories}, 2023.

\bibitem{gu2023investigating}
Y.~Gu, L.~Ying, Y.~Pu, X.~Hu, H.~Chai, R.~Wang, X.~Gao, and H.~Duan, ``Investigating package related security threats in software registries,'' in \emph{2023 IEEE Symposium on Security and Privacy (SP)}.\hskip 1em plus 0.5em minus 0.4em\relax IEEE, 2023, pp. 1578--1595.

\bibitem{datadog}
Guarddog, ``{Malicious Software Packages Dataset.}'' \url{https://github.com/datadog/malicious-software-packages-dataset}, 2023.

\bibitem{guo2023empirical}
W.~Guo, Z.~Xu, C.~Liu, C.~Huang, Y.~Fang, and Y.~Liu, ``An empirical study of malicious code in pypi ecosystem,'' in \emph{2023 38th IEEE/ACM International Conference on Automated Software Engineering (ASE)}.\hskip 1em plus 0.5em minus 0.4em\relax IEEE, 2023, pp. 166--177.

\bibitem{ruby_hust}
HUST, ``Hust rubygems mirror for downloads and installations,'' \url{https://mirrors.hust.edu.cn/rubygems/}.

\bibitem{ta577}
T.~Insight, ``{TA577’s Unusual Attack Chain Leads to NTLM Data Theft .}'' \url{https://www.proofpoint.com/us/blog/threat-insight/ta577s-unusual-attack-chain-leads-ntlm-data-theft}, 2024.

\bibitem{jang2011bitshred}
J.~Jang, D.~Brumley, and S.~Venkataraman, ``Bitshred: feature hashing malware for scalable triage and semantic analysis,'' in \emph{Proceedings of the 18th ACM conference on Computer and communications security}, 2011, pp. 309--320.

\bibitem{ssc_backdoor}
J.~Koljonen, ``{Warning! is rest-client 1.6.13 hijacked?}'' \url{https://github.com/rest-client/rest-client/issues/713}, 2019.

\bibitem{korczynski2017capturing}
D.~Korczynski and H.~Yin, ``Capturing malware propagations with code injections and code-reuse attacks,'' in \emph{Proceedings of the 2017 ACM SIGSAC Conference on Computer and Communications Security}, 2017, pp. 1691--1708.

\bibitem{ladisa2023sok}
P.~Ladisa, H.~Plate, M.~Martinez, and O.~Barais, ``Sok: Taxonomy of attacks on open-source software supply chains,'' in \emph{2023 IEEE Symposium on Security and Privacy (SP)}.\hskip 1em plus 0.5em minus 0.4em\relax IEEE, 2023, pp. 1509--1526.

\bibitem{lidisa2022javabytecode}
\BIBentryALTinterwordspacing
P.~Ladisa, H.~Plate, M.~Martinez, O.~Barais, and S.~E. Ponta, ``Towards the detection of malicious java packages,'' in \emph{Proceedings of the 2022 ACM Workshop on Software Supply Chain Offensive Research and Ecosystem Defenses}, ser. SCORED'22.\hskip 1em plus 0.5em minus 0.4em\relax New York, NY, USA: Association for Computing Machinery, 2022, pp. 63 -- 72. [Online]. Available: \url{https://doi.org/10.1145/3560835.3564548}
\BIBentrySTDinterwordspacing

\bibitem{langchain}
\BIBentryALTinterwordspacing
LangSmith, ``{LangChain, a unified platform for debugging, testing, evaluating, and monitoring your LLM applications.}'' 2023. [Online]. Available: \url{https://blog.langchain.dev/announcing-langsmith/}
\BIBentrySTDinterwordspacing

\bibitem{li2023hitchhiker}
H.~Li, Y.~Hao, Y.~Zhai, and Z.~Qian, ``The hitchhiker's guide to program analysis: A journey with large language models,'' \emph{arXiv preprint arXiv:2308.00245}, 2023.

\bibitem{liu2023languages}
H.~Liu, C.~Sferrazza, and P.~Abbeel, ``Languages are rewards: Hindsight finetuning using human feedback,'' \emph{arXiv preprint arXiv:2302.02676}, 2023.

\bibitem{ma2023scope}
W.~Ma, S.~Liu, W.~Wang, Q.~Hu, Y.~Liu, C.~Zhang, L.~Nie, and Y.~Liu, ``The scope of chatgpt in software engineering: A thorough investigation,'' \emph{arXiv preprint arXiv:2305.12138}, 2023.

\bibitem{gadget}
microsoft org., ``{OSS Gadget is a collection of tools that can help analyze open-source projects.}'' \url{https://github.com/microsoft/OSSGadget}, 2020.

\bibitem{backstabbers-online}
M.~Ohm, ``{Backstabber's Knife Collection},'' \url{https://dasfreak.github.io/Backstabbers-Knife-Collection/}, 2020.

\bibitem{Ohmssc2020}
M.~Ohm, H.~Plate, A.~Sykosch, and M.~Meier, ``Backstabber's knife collection: A review of open source software supply chain attacks,'' in \emph{Detection of Intrusions and Malware, and Vulnerability Assessment}, C.~Maurice, L.~Bilge, G.~Stringhini, and N.~Neves, Eds.\hskip 1em plus 0.5em minus 0.4em\relax Cham: Springer International Publishing, 2020, pp. 23--43.

\bibitem{ohm2020backstabber}
------, ``Backstabber’s knife collection: A review of open source software supply chain attacks,'' in \emph{International Conference on Detection of Intrusions and Malware, and Vulnerability Assessment}.\hskip 1em plus 0.5em minus 0.4em\relax Springer, 2020, pp. 23--43.

\bibitem{tongyi}
A.~Org., ``{Tongyi Qianwen 2.0 and Industry-specific Models to Support Customers of Generative AI.}'' \url{https://tongyi.aliyun.com/qianwen/}, 2023.

\bibitem{gemini}
G.~Org., ``{Google's Gemini family for the multi-modal model.}'' \url{https://poe.com/Gemini-Pro}, 2023.

\bibitem{faiss}
M.~org., ``Faiss (facebook ai similarity search) is a library that allows developers to quickly search for embeddings.'' \url{https://pytorch.org/}, 2017.

\bibitem{llama}
M.~Org., ``{Llama 2: open source, free for research and commercial use.}'' \url{https://llama.meta.com/llama2/}, 2023.

\bibitem{openai}
O.~Org., ``{The OpenAI API is used for a range of models and fine-tune custom models.}'' \url{https://platform.openai.com/docs/introduction}, 2023.

\bibitem{phylum-online}
P.~org., ``{The Software Supply Chain Security Company},'' \url{https://blog.phylum.io}, 2023.

\bibitem{tianwen-online}
Q.~A. org., ``{Tianwen: Software Supply Chain.}'' \url{https://tianwen.qianxin.com/home}, 2023.

\bibitem{snyk-online}
S.~org., ``{Snyk Security Database},'' \url{https://security.snyk.io/vuln}, 2023.

\bibitem{socket-online}
------, ``{Secure your supply chain. Ship with confidence.}'' \url{https://socket.dev}, 2023.

\bibitem{maloss}
Osssanitizer, ``{Maloss-sample},'' \url{https://github.com/osssanitizer/maloss-samples/}, 2020.

\bibitem{pearce2023examining}
H.~Pearce, B.~Tan, B.~Ahmad, R.~Karri, and B.~Dolan-Gavitt, ``Examining zero-shot vulnerability repair with large language models,'' in \emph{2023 IEEE Symposium on Security and Privacy (SP)}.\hskip 1em plus 0.5em minus 0.4em\relax IEEE, 2023, pp. 2339--2356.

\bibitem{pearce2022pop}
H.~Pearce, B.~Tan, P.~Krishnamurthy, F.~Khorrami, R.~Karri, and B.~Dolan-Gavitt, ``Pop quiz! can a large language model help with reverse engineering?'' \emph{arXiv preprint arXiv:2202.01142}, 2022.

\bibitem{pei2023can}
K.~Pei, D.~Bieber, K.~Shi, C.~Sutton, and P.~Yin, ``Can large language models reason about program invariants?'' in \emph{International Conference on Machine Learning}.\hskip 1em plus 0.5em minus 0.4em\relax PMLR, 2023, pp. 27\,496--27\,520.

\bibitem{qian2022malicious}
Y.~Qian, Y.~Zhang, N.~Chawla, Y.~Ye, and C.~Zhang, ``Malicious repositories detection with adversarial heterogeneous graph contrastive learning,'' in \emph{Proceedings of the 31st ACM International Conference on Information \& Knowledge Management}, 2022, pp. 1645--1654.

\bibitem{rokon2020sourcefinder}
M.~O.~F. Rokon, R.~Islam, A.~Darki, E.~E. Papalexakis, and M.~Faloutsos, ``$\{$SourceFinder$\}$: Finding malware $\{$Source-Code$\}$ from publicly available repositories in $\{$GitHub$\}$,'' in \emph{23rd International Symposium on Research in Attacks, Intrusions and Defenses (RAID 2020)}, 2020, pp. 149--163.

\bibitem{rossow2012prudent}
C.~Rossow, C.~J. Dietrich, C.~Grier, C.~Kreibich, V.~Paxson, N.~Pohlmann, H.~Bos, and M.~Van~Steen, ``Prudent practices for designing malware experiments: Status quo and outlook,'' in \emph{2012 IEEE symposium on security and privacy}.\hskip 1em plus 0.5em minus 0.4em\relax IEEE, 2012, pp. 65--79.

\bibitem{ssc_solar}
S.~M.~K. Saheed~Oladimeji, ``{SolarWinds hack explained: Everything you need to know.}'' \url{https://www.techtarget.com/whatis/feature/SolarWinds-hack-explained-Everything-you-need-to-know}, 2023.

\bibitem{SejiaAdria2022Machinelearn}
\BIBentryALTinterwordspacing
A.~Sejfia and M.~Sch\"{a}fer, ``Practical automated detection of malicious npm packages,'' in \emph{Proceedings of the 44th International Conference on Software Engineering}, ser. ICSE '22.\hskip 1em plus 0.5em minus 0.4em\relax New York, NY, USA: Association for Computing Machinery, 2022, pp. 1681 -- 1692. [Online]. Available: \url{https://doi.org/10.1145/3510003.3510104}
\BIBentrySTDinterwordspacing

\bibitem{sejfia2022practical}
A.~Sejfia and M.~Sch{\"a}fer, ``Practical automated detection of malicious npm packages,'' in \emph{Proceedings of the 44th International Conference on Software Engineering}, 2022, pp. 1681--1692.

\bibitem{semgrep}
semgrep org., ``{SemGrep rules for the security static analysis.}'' \url{https://github.com/semgrep/semgrep}, 2019.

\bibitem{ssc_fallguy}
A.~Sharma, ``{Inside the “Fallguys” Malware That Steals Your Browsing Data and Gaming IMs; Continued Attack on Open Source Software},'' \url{https://blog.sonatype.com/inside-the-fallguys-malware}, 2020.

\bibitem{shinn2023reflexion}
N.~Shinn, F.~Cassano, A.~Gopinath, K.~R. Narasimhan, and S.~Yao, ``Reflexion: Language agents with verbal reinforcement learning,'' in \emph{Thirty-seventh Conference on Neural Information Processing Systems}, 2023.

\bibitem{oss_report}
Sonatype, ``{State of the software supply chain},'' \url{https://www.sonatype.com/resources/state-of-the-software-supply-chain-2021}, 2021.

\bibitem{sun2023automatic}
W.~Sun, C.~Fang, Y.~You, Y.~Miao, Y.~Liu, Y.~Li, G.~Deng, S.~Huang, Y.~Chen, Q.~Zhang \emph{et~al.}, ``Automatic code summarization via chatgpt: How far are we?'' \emph{arXiv preprint arXiv:2305.12865}, 2023.

\bibitem{ruby_sysu}
SYSU, ``Sysu rubygems mirror for accelerated downloads,'' \url{http://mirror.sysu.edu.cn/rubygems/}.

\bibitem{taylor2020spellbound}
M.~Taylor, R.~K. Vaidya, D.~Davidson, L.~De~Carli, and V.~Rastogi, ``Spellbound: Defending against package typosquatting,'' \emph{arXiv preprint arXiv:2003.03471}, 2020.

\bibitem{pypi_tsinghua}
TUNA, ``Tuna pypi mirror for users in china,'' \url{https://pypi.tuna.tsinghua.edu.cn/}.

\bibitem{ruby_tsinghua}
------, ``Tuna rubygems mirror aiming to accelerate installations in china,'' \url{https://mirrors.tuna.tsinghua.edu.cn/rubygems/}.

\bibitem{SSC_Ledger}
I.~Turunen, ``{Decrypting the Ledger connect-kit compromise: A deep dive into the crypto drainer attack.}'' \url{https://www.sonatype.com/blog/decrypting-the-ledger-connect-kit-compromise-a-deep-dive-into-the-crypto-drainer-attack}, 2023.

\bibitem{ullah2023can}
S.~Ullah, M.~Han, S.~Pujar, H.~Pearce, A.~Coskun, and G.~Stringhini, ``Can large language models identify and reason about security vulnerabilities? not yet,'' \emph{arXiv preprint arXiv:2312.12575}, 2023.

\bibitem{vx_underground}
V.~Underground, ``{The largest collection of malware source code, samples, and papers on the internet.}'' url{https://www.vx-underground.org/}, 2007.

\bibitem{npm_ustc}
USTC, ``Ustc npm mirror for users in china,'' \url{https://mirrors.ustc.edu.cn/npm/}.

\bibitem{pypi_ustc}
------, ``Ustc pypi mirror for users in china,'' \url{https://pypi.mirrors.ustc.edu.cn/}.

\bibitem{vu2021lastpymile}
D.-L. Vu, F.~Massacci, I.~Pashchenko, H.~Plate, and A.~Sabetta, ``Lastpymile: identifying the discrepancy between sources and packages,'' in \emph{Proceedings of the 29th ACM Joint Meeting on European Software Engineering Conference and Symposium on the Foundations of Software Engineering}, 2021, pp. 780--792.

\bibitem{vu2023bad}
D.-L. Vu, Z.~Newman, and J.~S. Meyers, ``Bad snakes: Understanding and improving python package index malware scanning,'' in \emph{2023 IEEE/ACM 45th International Conference on Software Engineering (ICSE)}.\hskip 1em plus 0.5em minus 0.4em\relax IEEE, 2023, pp. 499--511.

\bibitem{vu2020typosquatting}
D.-L. Vu, I.~Pashchenko, F.~Massacci, H.~Plate, and A.~Sabetta, ``Typosquatting and combosquatting attacks on the python ecosystem,'' in \emph{2020 IEEE European Symposium on Security and Privacy Workshops (EuroS\&PW)}.\hskip 1em plus 0.5em minus 0.4em\relax IEEE, 2020, pp. 509--514.

\bibitem{guarddog}
E.~Wang, ``{The CLI tool that allows to identify malicious PyPI and npm packages},'' \url{https://github.com/DataDog/guarddog}, 2020.

\bibitem{wang2023voyager}
G.~Wang, Y.~Xie, Y.~Jiang, A.~Mandlekar, C.~Xiao, Y.~Zhu, L.~Fan, and A.~Anandkumar, ``Voyager: An open-ended embodied agent with large language models,'' \emph{arXiv preprint arXiv:2305.16291}, 2023.

\bibitem{wei2022chain}
J.~Wei, X.~Wang, D.~Schuurmans, M.~Bosma, F.~Xia, E.~Chi, Q.~V. Le, D.~Zhou \emph{et~al.}, ``Chain-of-thought prompting elicits reasoning in large language models,'' \emph{Advances in Neural Information Processing Systems}, vol.~35, pp. 24\,824--24\,837, 2022.

\bibitem{llmagent}
\BIBentryALTinterwordspacing
L.~Weng, ``{LLM-powered Autonomous Agents.}'' 2023. [Online]. Available: \url{lilianweng.github.io}
\BIBentrySTDinterwordspacing

\bibitem{wy2022InstalltimeAtt}
\BIBentryALTinterwordspacing
E.~Wyss, A.~Wittman, D.~Davidson, and L.~De~Carli, ``Wolf at the door: Preventing install-time attacks in npm with latch,'' in \emph{Proceedings of the 2022 ACM on Asia Conference on Computer and Communications Security}, ser. ASIA CCS '22.\hskip 1em plus 0.5em minus 0.4em\relax New York, NY, USA: Association for Computing Machinery, 2022, pp. 1139 -- 1153. [Online]. Available: \url{https://doi.org/10.1145/3488932.3523262}
\BIBentrySTDinterwordspacing

\bibitem{xia2023keep}
C.~S. Xia and L.~Zhang, ``Keep the conversation going: Fixing 162 out of 337 bugs for 0.42 each using chatgpt,'' \emph{arXiv preprint arXiv:2304.00385}, 2023.

\bibitem{xiao2021abusing}
F.~Xiao, J.~Huang, Y.~Xiong, G.~Yang, H.~Hu, G.~Gu, and W.~Lee, ``Abusing hidden properties to attack the node. js ecosystem,'' in \emph{30th USENIX Security Symposium (USENIX Security 21)}, 2021, pp. 2951--2968.

\bibitem{yao2023tree}
S.~Yao, D.~Yu, J.~Zhao, I.~Shafran, T.~L. Griffiths, Y.~Cao, and K.~Narasimhan, ``Tree of thoughts: Deliberate problem solving with large language models,'' \emph{arXiv preprint arXiv:2305.10601}, 2023.

\bibitem{yao2022react}
S.~Yao, J.~Zhao, D.~Yu, N.~Du, I.~Shafran, K.~Narasimhan, and Y.~Cao, ``React: Synergizing reasoning and acting in language models,'' \emph{arXiv preprint arXiv:2210.03629}, 2022.

\bibitem{zhang2020cyber}
Y.~Zhang, Y.~Fan, S.~Hou, Y.~Ye, X.~Xiao, P.~Li, C.~Shi, L.~Zhao, and S.~Xu, ``Cyber-guided deep neural network for malicious repository detection in github,'' in \emph{2020 IEEE International Conference on Knowledge Graph (ICKG)}.\hskip 1em plus 0.5em minus 0.4em\relax IEEE, 2020, pp. 458--465.

\bibitem{zimmermann2019small}
M.~Zimmermann, C.-A. Staicu, C.~Tenny, and M.~Pradel, ``Small world with high risks: A study of security threats in the npm ecosystem,'' in \emph{28th USENIX Security Symposium (USENIX Security 19)}, 2019, pp. 995--1010.

\bibitem{scrapy}
Zyte, ``{Scrapy, An open source and collaborative framework for extracting the data you need from websites.}'' \url{https://scrapy.org}, 2021.

\end{thebibliography}


\appendix

\begin{table}[htb] 
    \centering
    \caption{The summarization interpreted malware without package metadata.}
    \label{tab:missing:1}
    \fontsize{7}{8}\selectfont
    \begin{tabular}{ c c c | c c c }
    \hline
        Pkg. Name        & LoC        & \makecell[c]{Meta\\data}  & Pkg. Name        & LoC        & \makecell[c]{Meta\\data} \\\hline
        bytedtrace       &  13        &     No       &  xoloeduccelifz       &  1       &     No        \\
        common       &  181       &     No        & wxpay\_comm       &  130       &     No        \\
        distrib-0.1      &  43       &     No        & wxpayproto       &  130       &     No        \\
        evilpip       &  453       &     No       & virtualnv-0.1.1       &  39       &     No        \\
        fjkslfjsdgbjkbnjfdkq       &  4       &     No       & Users       &  14       &     No        \\
        management       &  0       &     No    & ttensorflow\_gpu       &  130       &     No        \\
        mgmtsdk\_v2       &  10801    &     No   & ttensorflow-gpu       &  130       &     No        \\
        mgmtsdk\_v2\_1       &  5042       &     No     & tequests       &  132       &     No        \\
        mumpy-0.1       &  39       &     No      & tdwTauthAuthentication    &  130       &     No   \\
        polaris       &  130       &     No      & SentinelOne.egg-info       &  0       &     No    \\
        pylibscate       &  1       &     No     & rfquests       &  132       &     No        \\
        PyLibUtil       &  16       &     No    &  requfsts       &  132       &     No        \\
        pypackscate       &  11       &     No   &  requestw       &  132       &     No        \\
        pypackscraper       &  11       &     No    &   requests       &  5471       &     No        \\
        r3quests       &  130       &     No    &    requeste       &  132       &     No        \\
        reeuests       &  132       &     No     & requesfs       &  132       &     No        \\ 
        \hline
    \end{tabular}%
\end{table}

\begin{table}[t] \small
    \centering
    \caption{ The detailed information of security analysis reports.}
    \label{tab:report:source:detailed}
    \begin{tabular}{c l   c }
    \hline
    Category                &   Website             &   \makecell[c]{Report \\  Num.\#   }         \\ \hline
    \multirow{16}{*}{\makecell[c]{Technical \\Community} }    &      github.com     &       52     \\ 
                            &                  techradar.com/news               &     40        \\ 
                            &                           zdnet.com/article        &    40         \\ 
                            &                           lore.kernel.org/all              &    38         \\ 
                            &                            forums.macrumors.com/thread    &      37       \\ 
                            &                            thenewstack.io    &       37      \\ 
                            &                            securelist.com     &       36      \\ 
                            &                            welivesecurity.com    &      36       \\ 
                            &                            research.checkpoint.com     &     34        \\ 
                            &                             lwn.net/Articles     &       32      \\ 
                            &                           theregister.com    &      32       \\ 
                            &                           securitylab.github.com/advisories    &    31          \\ 
                            &                           static.lwn.net/kerneldoc    &      30       \\ 
                            &                           cn-sec.com    &      19       \\ 
                            &                           bleepingcomputer.com    &       12      \\ 
                            &                           forum.portswigger.net/thread        &  10    \\ \hline
    \multirow{14}{*}{\makecell[c]{Commercial\\ org.} }  &      blog.sonatype.com               &         86    \\ 
                            &                                sonatype.com               &         69   \\
                            &                        blog.avast.com  &        38     \\ 
                            &                       microsoft.com/en-us/security/blog    &      37       \\ 
                            &                      unit42.paloaltonetworks.com   &      37       \\ 
                            &                       reversinglabs.com/blog   &          32\\ 
                            &                       blog.phylum.io    &        31     \\ 
                            &                       sonarsource.com/blog   &      31       \\ 
                            &                       snyk.io/blog    &       30      \\ 
                            &                       blog.npmjs.org/post     &     28        \\ 
                            &                       vulcan.io/blog    &      28       \\ 
                            &                       mend.io/blog   &           27  \\ 
                            &                       jfrog.com/blog   &           25  \\ 
                            &                       checkmarx.com/blog   &         23    \\ 
                            &                       qianxin.com/blog             &  23    \\\hline
    \multirow{4}{*}{News }         &       thehackernews.com           &      44       \\ 
                                    &                      krebsonsecurity.com   &     37        \\ 
                                    &                      therecord.media   &        33     \\ 
                                    &                      portswigger.net/daily-swig        &  29    \\\hline
    \multirow{3}{*}{Individual }    &      medium.com          &   46 \\
                                &                           bertusk.medium.com             & 12 \\   
                                &                           twitter.com             & 37  \\\hline
    \multirow{1}{*}{Official }     &     nbu.gov.sk              & 24   \\  \hline
    \multirow{1}{*}{Other }    &        29   websites        & 43   \\  
    \hline
    \end{tabular}%
\end{table}

\begin{table}[!htbp] \small
    \centering
    \caption{The summarization of interpreted malware with more than 15,000 LoC.}
    \label{tab:missing:2}
    \begin{tabular}{ l r  }
    \hline
        Pkg. Name                    & LoC                  \\\hline
    \makecell[l]{awscl-1.27.67, awsclie-1.27.67, awsclii-1.27.67}    &  38,320              \\
        apidev-coop-1.2.26       &  32,128              \\
        pythonkafka-1.3.5       &  25,457            \\
     \makecell[l]{  urlib3-1.21.1, urllib-1.21.1 }      &  20,448              \\
      \makecell[l]{s3tranfer-0.5.2, s3transfere-0.5.2, s3transferr-0.5.2 \\
      s3trnasfer-0.5.2, s3trnasfers-0.5.2}      &  19,614            \\
  \makecell[l]{ python-dateutils-2.9.9, python-dateutils-3.0.0, 3.0.1\\
  python-dateutils-3.0.2,  python-dateutils-3.0.3}      &  18933            \\
        flsak-2.2.3       &  18255            \\
      \makecell[l]{ liblxml3-1.5, liblxml3-0.2 , liblxml3-1.4 \\
      liblxml3-1.7, liblxml3-0.1m, liblxml3-2.0 \\
      liblxml3-2.1, liblxml3-2.2, liblxml3-2.2\\
      liblxml3-2.6,  liblxml3-2.7}         &  16512           \\
     \makecell[l]{mirrorbot-1.3, mirrorbot-0.9, mirrorbot-1.1}      &  16,507           \\
 \makecell[l]{sentinelone-1.1.7,  sentinelone-1.1.8, 1.1.9,  \\
  sentinelone-1.2.0, sentinelone-1.2.1, 1.1.6,  \\ 
  sentinelone-1.1.3, sentinelone-1.1.4, 1.1.5,  \\
  sentinelone-1.0.9, sentinelone-1.1.0, 1.0.7, \\
   sentinelone-1.0.8, sentinelone-1.0.3,  1.0.4, \\
    sentinelone-1.0.5,     sentinelone-1.1.2, 1.0.6, \\
    sentinelone-1.0.1, sentinelone-1.0.2,  1.0.0    }      &  \makecell[r]{16,\\000+}          \\
        sentinelonesdk-1.0.0       &  15,332         \\

        \hline
    \end{tabular}%
\end{table}

\begin{table}[h] \small
    \centering
    \caption{The summarization of interpreted malware with 0 LoC.}
    \label{tab:missing:3}
    \begin{tabular}{ l r  }
    \hline
        Pkg. Name                    & LoC                  \\\hline
        broke-rcl-0.0.0       &  0             \\
\makecell[l]{    bytedtrace-0.1.10.dist-info, bytedtrace-0.1.11.dist-info \\
bytedtrace-0.1.12.dist-info,  bytedtrace-0.1.13.dist-info\\
bytedtrace-0.1.14.dist-info,  bytedtrace-0.1.15.dist-info\\
bytedtrace-0.1.16.dist-info, bytedtrace-0.1.17.dist-info \\
bytedtrace-0.1.18.dist-info,  bytedtrace-0.1.19.dist-info \\
bytedtrace-0.1.2.dist-info,  bytedtrace-0.1.20.dist-info  \\ 
bytedtrace-0.1.6.dist-info,  bytedtrace-0.1.7.dist-info \\ 
bytedtrace-0.1.8.dist-info,  bytedtrace-0.10.1.dist-info \\ 
bytedtrace-0.10.2.dist-info,  bytedtrace-0.10.3.dist-info  \\ 
bytedtrace-0.10.4.dist-info,  bytedtrace-0.2.4.dist-info \\
bytedtrace-0.2.5.dist-info,  bytedtrace-0.2.6.dist-info \\
bytedtrace-0.2.7.dist-info,  bytedtrace-0.2.8.dist-info \\
bytedtrace-0.3.0.dist-info,  bytedtrace-0.3.1.dist-info \\
bytedtrace-0.3.10.dist-info,  bytedtrace-0.3.2.dist-info \\
bytedtrace-0.3.3.dist-info,  bytedtrace-0.3.4.dist-info \\
bytedtrace-0.3.5.dist-info,  bytedtrace-0.3.6.dist-info \\
bytedtrace-0.3.7.dist-info,  bytedtrace-0.3.8.dist-info \\
bytedtrace-0.3.9.dist-info  }       &  0            \\
        fjkslfjsdgbjkbnjfdkq-1.1.1.dist-info       &  0              \\
        h8shdf89d-2.28.2.dist-info       &  0             \\
        management       &  0            \\
   \makecell[l]{ otr-utils-4.1.0, otr-utils-4.1.9}       &  0              \\
        packagescraperlib-1.4.0.dist-info       &  0             \\
\makecell[l]{   pylibscate-1.1.0.dist-info, pylibscate-1.1.1.dist-info}       &  0     \\

 \makecell[l]{pylibutil-0.9.2.dist-info,  pylibutil-0.9.3.dist-info}   &  0            \\
        \makecell[l]{   pypackscate-1.1.1.dist-info, pypackscate-1.1.3.dist-info \\
        pypackscate-1.2.0.dist-info,  pypackscate-1.3.0.dist-info\\
        pypackscraper-1.0.4.dist-info, 1.0.5.dist-info  \\
        pypackscraper-1.0.6.dist-info}     &  0              \\
\makecell[l]{r3quests-2.28.1.dist-info, reeuests-2.28.1.dist-info \\
requesfs-2.28.1.dist-info, requeste-2.28.1.dist-info\\
requestw-2.28.1.dist-info, requfsts-2.28.1.dist-info\\
rfquests-2.28.1.dist-info   }     &  0            \\
        SentinelOne.egg-info       &  0              \\
        tequests-2.28.1.dist-info       &  0              \\
        test23414234234-0.5       &  0            \\
        xiedemo-0.0.0.dist-info       &  0             \\
        xoloeduccelifz-0.0.0.dist-info       &  0           \\
        \hline
    \end{tabular}%
\end{table}

\begin{table*}[hb]
\centering
\caption{The prompt for {\tech}. The \emph{Guidelines} provide a structured approach to analyze malware packages for security threats. \textcolor{blue}{Blue}/\textcolor{orange}{Orange} indicates the user-provided paths and relevant information.}
\label{tab:prompt}
\begin{tabular}{p{16.5cm}}
\\ \toprule
\textbf{Executor} \\
\midrule
You are an Analysis Agent responsible for performing a context-based analysis to identify potential security threats in a Python package.

Your goal is to perform a detailed analysis to identify potential security threats using the provided \textcolor{orange}{\{context\}}, which includes names and descriptions of attack vectors. Follow these specified steps and guidelines:

\textbf{Guidelines:}
\begin{enumerate}
    \item \textbf{Step 1: Input Information}
    \begin{itemize}
        \item \textbf{Input:} User provides package metadata and Python code as \textcolor{orange}{\{file\_content\}}.
        \item \textbf{Output:} Catalog of Python files and their respective execution sequences.
    \end{itemize}
    
    \item \textbf{Step 2: Detection of Deceptive AV Attack Vectors in pkg-info}
    \begin{itemize}
        \item \textbf{Context:} Utilize the user-provided \textcolor{orange}{\{file\_content\}} to identify specific deceiving attack vectors (names and descriptions)\textcolor{orange}{\{context\}}.
        \item \textbf{Action:} Examine pkg-info for characteristics matching the descriptions of deceiving attack vectors in the context, focusing on information that is suspicious, unknown, or uncertain.
        \item \textbf{Output:} List of identified deceiving attack vectors based on context.
    \end{itemize}
    
    \item \textbf{Step 3: Detection of Execution AV Attack Vectors in Python Files}
    \begin{itemize}
        \item \textbf{Context:} Use the execution sequence from Step 1 and the \textcolor{orange}{\{context\}} for targeted analysis.
        \item \textbf{Action:} Scrutinize Python files for potential malicious attack vectors, focusing on their execution sequence and matching against \textcolor{orange}{\{context\}}.
        \item \textbf{Output:} List of identified malicious attack vectors based on \textcolor{orange}{\{context\}} with the execution sequence.
    \end{itemize}
    
    \item \textbf{Step 4: Output the Names of Attack Vectors and the TTPs}
    \begin{itemize}
        \item \textbf{Action:} Compile the findings from Steps 2 and 3. Focus on reporting the names of identified attack vectors.
        \item \textbf{Output:} A list that includes names of all identified vectors (deceiving and execution).Use the ``$\rightarrow$" symbol to link the attack vectors of the python packages.
    \end{itemize}
\end{enumerate}

\textbf{Note that:} Progress through each step methodically, ensuring a thorough examination based on the provided context. Document findings and observations systematically at each step. Continuously cross-reference with context to ensure alignment with specified attack vector descriptions. Provide regular summaries to maintain clarity and focus on the objectives.

Now, you will be provided with the following information; please generate your response according to this information:

\textcolor{orange}{Package metadata and python code: \{file\_content\}}

\textcolor{orange}{Information of attack vectors: \{context\}}

\textcolor{blue}{System Message: Ensure operations are always completed in user-specified directories.}
\\
\bottomrule
\end{tabular}

\end{table*}

\begin{table*}[!t]
\centering
\caption{The prompt for an LLM-based agent. The \emph{Guidelines} provide a structured approach to decompress zip files. \textcolor{blue}{Blue}/\textcolor{orange}{Orange} indicates the user-provided paths and relevant information.}
\label{tab:genttp:agent}
\begin{tabular}{p{16.5cm}}
\\ \toprule
\textbf{Executor} \\
\midrule
You are an Unzip Agent responsible for decompressing zip files from a specified directory to another specified directory.

Your goal is to ensure that all zip files are properly decompressed following the specified steps and guidelines.

You should follow these guidelines:

1. \textbf{Set the Working Directory:} Ensure the current directory is set to the specified raw path: \textcolor{orange}{\{raw\_path\}}. If not, change the directory using ``cd \textcolor{orange}{\{raw\_path\}}''.

2. \textbf{Decompress Zip Files:} Decompress all the zip files from the specified path: \textcolor{orange}{\{raw\_path\}} to \textcolor{orange}{\{unzip\_path\}}. 

3. \textbf{Shell Commands:} Use Python scripts or shell commands to complete these tasks.

4. \textbf{Errors:} If errors occur, log them and try alternative methods.

\textbf{Note that:} Each extracted file should be saved in a newly created directory corresponding to the zip filename to prevent errors during subsequent decompressions of files with similar names. All paths should be handled using UTF-8 encoding.

Now, you will be provided with the following information; please generate your response according to this information:

\textcolor{orange}{Raw Path: \{raw\_path\}}

\textcolor{orange}{Unzip Path: \{unzip\_path\}}

\textcolor{blue}{System Message: Ensure operations are always completed in user-specified directories.}
\\
\bottomrule
\end{tabular}

\end{table*}

\begin{table*}[!t]
\centering
\caption{The prompt for analysis reports. The \emph{Guidelines} provide a structured approach to identify malicious Python packages from URLs. \textcolor{blue}{Blue}/\textcolor{orange}{Orange} indicates the user-provided paths and relevant information.}
\label{tab:prompt:report}
\begin{tabular}{p{16.5cm}}
\\ \toprule
\textbf{Executor} \\
\midrule
You are an Analysis Agent responsible for identifying malicious Python packages from a list of URLs.

Your goal is to extract the names and key actions of malicious PyPI packages from the provided URLs. Follow these specified steps and guidelines:

\textbf{Guidelines:}
\begin{enumerate}
    \item \textbf{Step 1: Get Website Sources}
    \begin{itemize}
        \item \textbf{Input:} User enters the required website keywords, including software supply chain, malicious conduct and malware packages.
        \item \textbf{Action:} Search for websites with relevant keywords.
        \item \textbf{Output:} Save the website in \textcolor{orange}{\{URL\_excel\}}.
    \end{itemize}

    \item \textbf{Step 2: Extract Information from URLs}
    \begin{itemize}
        \item \textbf{Input:} Use the extraction \textcolor{orange}{\{schema\}} definition as ["malicious\_package\_name", "malicious\_package\_actions"].
        \item \textbf{Action:} Extract the name of the malicious pypi package present in \textcolor{orange}{\{URL\_excel\}} and the key description.
        \item \textbf{Output:} Extracted information including package names and actions.
    \end{itemize}
    
    \item \textbf{Step 4: Save Extracted Information}
    \begin{itemize}
        \item \textbf{Action:} Save the extracted information to an Excel file.
        \item \textbf{Output:} Excel file containing the names and actions of identified malicious packages.
    \end{itemize}
\end{enumerate}

\textbf{Note that:} Progress through each step methodically, ensuring a thorough examination based on the provided \textcolor{orange}{\{schema\}}. Provide regular summaries to maintain clarity and focus on the objectives.

Now you will be provided with the following information, please generate your response according to this information:

\textcolor{orange}{Website: \{URL\_excel\}}

\textcolor{orange}{Extraction Schema: \{schema\}}

\textcolor{blue}{System Message: Ensure operations are always completed in user-specified directories.}
\\
\bottomrule
\end{tabular}

\end{table*}

\begin{table}[hb]
\centering
\caption{The prompt for ChatBot. The \emph{Guidelines} provide a structured approach to generate responses. \textcolor{blue}{Blue}/\textcolor{orange}{Orange} indicates the user-provided question and relevant information.}
\label{tab:chatbot}
\begin{tabular}{p{8.0cm}}
\\ \toprule
\textbf{Executor} \\
\midrule
You are a Malware Package Behavior Analysis Engineer working on software supply chain security. Your task is to generate a response based on the user's \textcolor{orange}{\{question\}} and relevant information retrieved from the \textcolor{orange}{\{database\}}.

TTP (Tactics, Techniques, and Procedures) is used to describe the attack behaviors and sequences of open source software malware. Specifically, TTPs include strategies (tactics) for tricking users into installing malware, and specific techniques (technologies and procedures) for executing malicious activities. These represent attack patterns for malware packages. Each JSON file in the \textcolor{orange}{\{database\}} contains information about a malware package, such as the package name, TTPs, and the analysis process of these TTPs.

Your goal is to provide a detailed and accurate response by referring to the ``analysis\_process" in the \textcolor{orange}{\{database\}} when explaining the TTP (Tactics, Techniques, and Procedures) analysis process for a malware package. Use the ``$\rightarrow$" symbol to link the ``TTP" in the \textcolor{orange}{\{database\}} for a malware package.

You should follow these guidelines:

1. \textbf{Factual Questions:} If the \textcolor{orange}{\{question\}} is factual, provide a direct and concise answer.

2. \textbf{Open-ended Questions:} If the \textcolor{orange}{\{question\}} is open-ended, offer a detailed and thoughtful response.

3. \textbf{Insufficient Information:} If the information is insufficient, acknowledge it and suggest possible next steps.

Now, you will be provided with the following information; please generate your response according to this information:

\textcolor{orange}{User Question: \{question\}}

\textcolor{orange}{Database: \{my\_custom\_collection\}}

\textcolor{blue}{System Message: Ensure operations are always completed in user-specified directories.}
\\
\bottomrule
\end{tabular}
\end{table}

\begin{table}[htbp]\small
    \centering
    \caption{The context of deceptive and execution TTPs. }
    \label{tab:context}
    \begin{tabular}{l | r  }
    \hline
    Term                   &   Description       \\ \hline  
    $typosquatting$          &  \makecell[r]{ a similar package name\\ to imitate a legitimate package}    \\  \hline    
    \makecell[l]{$imitated$-\\$version$ }        &  \makecell[r]{ uses a common version number \\ as a legitimate version }   \\      \hline 
    \makecell[l]{$fake$-\\$description$ }&  \makecell[r]{ a common description \\ imitating  a legitimate package description}    \\   \hline 
    \makecell[l]{$malicious$-\\$dependence$}    &  \makecell[r]{  uses a different malware library \\ as its dependency}    \\    \hline 
    \makecell[l]{$imitated$-\\$url$ }             &    \makecell[r]{ imitates a legitimate \\ URL for its homepage } \\  \hline 
    \makecell[l]{$fake$-\\$contact$}           & \makecell[r]{  a contact to imitate authors \\ in a legitimate package}     \\     \hline   
    $pre$-$install$                      &   a pre-install script running commands     \\  \hline 
$install$     & an install script running commands   \\\hline 
$post$-$install$                      &   a post-install script running commands     \\  \hline 
$cmd$                       &    \makecell[r]{execution command: \\ setup, exec, getoutput,  call \\  check\_output, run, eval, popen  } \\\hline 
$evasion$                     & \makecell[r]{obscuring function: \\ base64, zlib.decompress}      \\  \hline 
$conceal$                     & \makecell[r]{concealing information:\\ exec, reveal, decode} \\  \hline 
$data$                     &    sensitive data \\  \hline 
$file$ / $dir$                      &    file/directory-related operation  \\  \hline 
$permission$                              &     changing permission of file, dir           \\ \hline 
$read$ / $write$                     &    read/write operation      \\   \hline 
$URL$ / $IP$ /$port$                   &    malicious URL,  IP address, port       \\   \hline 
$send$ /  $recieve$                   &    sending or receive data       \\  
\hline
\end{tabular}%

\end{table}

\end{sloppypar}
\end{document}